\documentclass[twocolumn,times,tighten]{aastex63}
\usepackage{hyperref}
\usepackage{mleftright}

\submitjournal{PASP}
\accepted{February 23, 2024}

\shorttitle{The ArgusSpec Prototype}
\shortauthors{Galliher et al.}

\begin{document}

\title{The ArgusSpec Prototype: Autonomous Spectroscopic Follow-up of Flares Detected by Large Array Telescopes}

\correspondingauthor{Nathan W. Galliher}
\email{nathan.galliher@unc.edu}

\author[0000-0001-8105-1042]{Nathan W. Galliher}
\affil{Department of Physics and Astronomy, University of North Carolina at Chapel Hill, 120 E. Cameron Ave., Chapel Hill, NC 27514, USA}

\author[0009-0005-7319-7355]{Thomas Procter}
\affil{Department of Physics and Astronomy, University of North Carolina at Chapel Hill, 120 E. Cameron Ave., Chapel Hill, NC 27514, USA}

\author[0000-0001-9380-6457]{Nicholas M. Law}
\affil{Department of Physics and Astronomy, University of North Carolina at Chapel Hill, 120 E. Cameron Ave., Chapel Hill, NC 27514, USA}

\author[0000-0002-6339-6706]{Hank Corbett}
\altaffiliation{National Science Foundation Graduate Research Fellow}
\affil{Department of Physics and Astronomy, University of North Carolina at Chapel Hill, 120 E. Cameron Ave., Chapel Hill, NC 27514, USA}

\author[0000-0002-0583-0949]{Ward S. Howard}
\affil{Department of Astrophysical and Planetary Sciences, University of Colorado, 2000 Colorado Avenue, Boulder, CO 80309, USA}
\affil{NASA Sagan Fellow}

\author[0000-0002-1906-1167]{Alan Vasquez Soto}
\affil{Department of Physics and Astronomy, University of North Carolina at Chapel Hill, 120 E. Cameron Ave., Chapel Hill, NC 27514, USA}

\author{Ramses Gonzalez}
\affil{Department of Physics and Astronomy, University of North Carolina at Chapel Hill, 120 E. Cameron Ave., Chapel Hill, NC 27514, USA}

\author[0009-0008-3299-3648]{Lawrence Machia}
\affil{Department of Physics and Astronomy, University of North Carolina at Chapel Hill, 120 E. Cameron Ave., Chapel Hill, NC 27514, USA}

\author[0000-0001-8544-584X]{Jonathan Carney}
\affil{Department of Physics and Astronomy, University of North Carolina at Chapel Hill, 120 E. Cameron Ave., Chapel Hill, NC 27514, USA}

\author[0009-0008-2179-8943]{William J. Marshall}
\affil{Department of Physics and Astronomy, University of North Carolina at Chapel Hill, 120 E. Cameron Ave., Chapel Hill, NC 27514, USA}

\begin{abstract}
ArgusSpec is a prototype autonomous spectroscopic follow-up instrument designed to characterize flares detected by the Argus Pathfinder telescope array by taking short exposure (30~s) broadband spectra (370 - 750~nm) at low resolutions (R$\sim$150 at 500~nm). The instrument is built from consumer off-the-shelf astronomical equipment, assembled inside a shipping container, and deployed alongside the Argus Pathfinder at a dark sky observing site in Western North Carolina. In this paper, we describe the hardware design, system electronics, custom control software suite, automated target acquisition procedure, and data reduction pipeline. We present initial on-sky test data used to evaluate system performance and show a series of spectra taken of a small flare from AD Leonis. The $\rm\$35k$ prototype ArgusSpec was designed, built, and deployed in under a year, largely from existing parts, and has been operating on-sky since March 2023. With current hardware and software, the system is capable of receiving an observation, slewing, performing autonomous slit acquisition, and beginning data acquisition within an average of 32 s. With Argus Pathfinder's 1-second-cadence survey reporting alerts of rising sources within 2 s of onset, ArgusSpec can reach new targets well within a minute of the start of the event. As built, ArgusSpec can observe targets down to a 20$\sigma$ limiting magnitude of $m_V \approx 13$ at 30 s cadence with an optical resolution of R$\sim$150 (at 500 nm). With automated rapid acquisition demonstrated, later hardware upgrades will be based on a clean-sheet optical design, solving many issues in the current system, significantly improving the limiting magnitude, and potentially enabling deep spectroscopy by the coaddition of data from an array of ArgusSpec systems. The primary science driver behind ArgusSpec is the characterization of the blackbody evolution of flares from nearby M-dwarfs. Large flares emitted by these stars could have significant impacts on the potential habitability of any orbiting exoplanets, but our current understanding of these events is in large part built on observations from a handful of active stars. ArgusSpec will characterize large numbers of flares from across the night sky, building a spectroscopic library of the most extreme events from a wide variety of stellar masses and ages. 
\end{abstract}



\section{Introduction \label{sec:introduction} } 

Stellar flares are highly energetic magnetic re-connection events that produce radiation across the electromagnetic spectrum, evolve on timescales of seconds to hours~\citep{2013ApJS..207...15K, 2015ApJ...809..104A}, and make up the majority of short timescale (minute-to-hour) galactic transients~\citep{2006ApJ...644L..63K}. Typical M-dwarf flares release between $10^{29}$ to $10^{32}$ erg while the largest flares, known as superflares, have energies over $10^{33}$~erg. Our current understanding of the spectroscopic characteristics of stellar flares and superflares is largely derived from observations of a limited number of active M-dwarfs, such as AD Leonis and EV Lacertae. These observations have set the canonical value for the effective temperatures of flares at 9,000 K~\citep{1991ApJ...378..725H, 2003ApJ...597..535H, 2013ApJS..207...15K}, which determines the energy budget of ultraviolet (UV) and optical emission \citep{2015ApJ...809...79O}.

The small number of simultaneous near-UV (NUV) and optical flare spectra that have been fit with blackbody models (e.g. \citealt{1991ApJ...378..725H, Robinson:1993, Kowalski:2019, Kowalski:2022}) suggest that improvements to the canonical model are needed. \citet{Kowalski:2019} find the NUV emission of two flares from GJ 1243 is 2-3$\times$ higher than predicted from a 9000 K blackbody fit to the optical spectra. Likewise, flare rates of field age M-dwarfs observed by GALEX and TESS show that using a 9000 K blackbody to scale the flare energy from the optical under-estimates the NUV by a factor of 6.5$\pm$0.7 \citep{Jackman:2023}. Simultaneous flare observations with GALEX and \textit{Kepler} find effective temperatures above 36,000 K are necessary to fit the NUV and optical observations \citep{Brasseur:2023}. 

Effective temperatures of the largest M-dwarf flares are particularly uncertain. The few superflares that have been observed spectroscopically show evidence for temperatures as high as 40,000 K~\citep{2005ApJ...633..447R, 2018ApJ...852...61K, 2019ApJ...871L..26F,Brasseur:2023}.~\cite{2020ApJ...902..115H} used simultaneous multi-wavelength observations from the Evryscope and TESS to estimate blackbody temperatures for 42 superflares and found that 5\% of the sample had estimated effective temperatures during the flare peak of over 30,000 K, with the largest flare reaching an estimated peak temperature of 42,000 K. Radiative-hydrodynamic modeling of the \citet{1991ApJ...378..725H} AD Leo superflare show the NUV is under-estimated by a factor of 2.6 when scaling from the red-optical using a 9000 K blackbody \citep{Kowalski:2022}. The models predict UV-C fluxes of 800-1000 W m$^{-2}$ reach temperate rocky planets during 10$^{34}$ erg superflares, consistent with the range estimated from broadband observations in \citet{2020ApJ...902..115H}. However, spectral observations of flares are required to demonstrate whether such high temperatures truly arise from the continuum or are due to contamination from lines.

\subsection{Flare Impacts on Exoplanet Habitability}
Rocky planets orbiting in the temperate zones of M-dwarfs typically receive relatively small amounts of UV flux from their host star in its quiescent state. Flares from the host star might provide enough excess radiation to support the emergence of prebiotic chemistry that would not arise otherwise due to a lack of UV flux~\citep{2017ApJ...843..110R, 2018SciA....4.3302R}. On the other hand, the high energy radiation from flares drives photochemical escape and photochemistry. Extreme-UV radiation from flares is capable of inducing significant hydrodynamic loss in planetary atmospheres \citep{France:2020}, while far-UV (FUV) and NUV radiation drives dis-equilibrium photochemical processes \citep{Loyd:2018, Ranjan:2020}.
Dis-equilibrium chemistry from flares and any accompanying high-energy particle emissions are capable of depleting any Earth-like atmosphere's ozone column depth by up to 94\% over time~\citep{2010AsBio..10..751S}. UV radiation from subsequent flares could then potentially sterilize the planet's surface~\citep{2018ApJ...867...70L, 2018ApJ...860L..30H, 2019AsBio..19...64T}. 

The flare rates of TESS Objects of Interest exoplanet host stars from the Transiting Exoplanet Survey Satellite are typically measured in the optical \citep{2022MNRAS.512L..60H}. Photochemical models therefore scale the optical energies of flares into the UV using the canonical 9000 K blackbody (e.g. \citealt{tilley2019modeling, Chen:2021, doAmaral_2022}). However, \citet{Kowalski:2019} find the canonical 9000 K to underestimate NUV emission by a factor of 2-3x. Flare temperature is, therefore, a key input in modeling the effects of stellar flaring on ozone depletion in the atmospheres of rocky habitable-zone exoplanets orbiting active M-dwarfs~\citep{2018ApJ...860L..30H, 2019AsBio..19...64T}. Furthermore, flares are known to have different spectral evolution as a function of their impulsiveness, which tracks the rate at which dominant flare emission moves from the photosphere up into the chromosphere and corona~\citep{kowalski2018evolution}. Because superflares are rare and stochastic, the relationship between flare temperature-color evolution and impulsiveness is not well studied. As an added complexity, the relative contributions from its stellar continuum and line emission, and the blackbody temperature itself, change throughout the flare~\citep{2013ApJS..207...15K}.

It is possible that the current canonical temperature values and blackbody behavior do not hold for a large unbiased sample of flares and therefore lead to incorrect model inputs for the UV flux. To correct for the biases in obtaining spectra from only a handful of exceptionally active stars, a  comprehensive spectroscopic survey of flares and superflares from a wide range of M-dwarfs is necessary to better constrain temperatures and UV fractional fluxes and therefore better constrain the modeling of their impacts on exoplanets. Building a large spectroscopic library comprised of hundreds of flares requires automated spectroscopic follow-up. 

\subsection{The ArgusSpec Targets: Detecting Flares and Other Transients with Argus Pathfinder}

The Argus Array~\citep{2022PASP..134c5003L} is a planned next-generation all-sky survey that is currently undergoing prototyping via the Argus Pathfinder Array~\citep{2022SPIE12182E..4HL}. Argus Pathfinder will survey the northern sky in stripes, with its 20 cm telescopes observing a 2.3 GPix field, 2.5° wide in RA and 88° long in declination. Each field will be observed at high cadence for 15 minutes each night. Pathfinder will pick between 30 s ("normal") and 1 s ("high-speed") exposure times to remain background-limited for as long as possible. Given a night's expected brightness, it will be designated into one of these two modes. It is expected that Pathfinder will spend 20-30\% of its observation time in the high-speed mode. The 1 s exposure time in the high-speed flare follow-up mode allows for the observation of a flare’s impulsive phase. While in this high-speed mode, Pathfinder is expected to have a bright limit of $m_g \sim 7$ and a limiting magnitude of $m_g \sim 16$. 

Pathfinder transient detections are made and reported in real-time by the Argus Array Hierarchical Data Processing System pipeline \citep[Argus-HDPS]{2022SPIE12189E..10C}, using core algorithms developed in the Evryscope Fast Transient Engine pipeline \citep[EFTE]{2023ApJS..265...63C}. Argus-HDPS produces a real-time event stream of candidate events, which can then be filtered based on cross-matching with reference catalogs, number of detections, magnitude, and various per-detection quality metrics. Incoming data is analyzed on a GPU cluster, where standard sensor calibrations are applied and the image is reprojected into a HEALPix-based representation for subsequent image subtraction. Multiple image-subtraction modes are available, depending on the necessary data throughput. In the one-second-cadence mode, where Pathfinder's data rate is 37 Gbps, incoming images are normalized based on their noise level and subtracted from a previous normalized image from the same pointing, separated in time by minutes. Candidates are identified in this ``discovery'' image, which has units of standard deviations. High-probability candidates, as determined by the machine-learning-based vetting system, are passed directly to ArgusSpec. This system is described in detail in \citep{2022SPIE12189E..10C, 2023ApJS..265...63C}. The core Argus-HDPS transient-detection algorithms have been validated in long-term and high-cadence use on the Evryscopes with EFTE. This system has been successful at detecting large numbers of flares and superflares (e.g., \citealt{2018ApJ...860L..30H, 2019ApJ...881....9H, 2020ApJ...902..115H, 2020ApJ...895..140H, 2020ApJ...900...27G}) along with satellite glints~\citep{2020ApJ...903L..27C} and more distant astrophysical transients (e.g., \citealt{2018ATel11467....1C}, \citealt{2020ApJ...899..162W}, \citealt{2021RNAAS...5..160Q}, \citealt{2023MNRAS.521.5453S}). On a typical observing night, the Evryscopes produce hundreds of high-confidence transient alerts, including around 10 detections of flares and superflares. Given its smaller field of view but superior depth, Argus Pathfinder, along with Argus-HDPS, is expected to return a similar rate of $10 \pm 5$ M-dwarf type flares per night. This rate is estimated using the number of M-dwarfs of each spectral type above the limiting magnitude and within the field of view, active fractions, average flare rates, and the fraction of nights with good weather. At this rate, ArgusSpec will be able to follow the majority of alerts without the need to categorize and prioritize superflares prior to beginning observations.

The real-time transient alert streams produced by Argus Pathfinder will provide an opportunity to build a spectroscopic library of superflares from across the sky by sending relevant detections to low-latency follow-up facilities. Until now, implementing a large-scale rapid flare follow-up program would have been challenging due to the difficulty of rapid triggering and the relatively low number of existing automated follow-up facilities. The number of impulsive phases observed per night will add significantly to the flare physics community’s pool of data as ArgusSpec continues to operate, most importantly, allowing a large-sample measurement of the temperatures of the brightest superflares at peak without the line contamination introduced by wideband photometry \citep{2020ApJ...902..115H}. By following the brightest events across the entire sky, far more extreme events will be observed than is reasonable to achieve from single-target staring campaigns and over a much larger variety of stellar ages and masses. In the following sections, we discuss some of the existing automated follow-up facilities to provide further context behind the construction of a new automated follow-up instrument.

\subsection{Existing Automated Follow-up Instruments}
Traditionally, automated slit acquisition has been seen as a major hurdle in the development of robotic spectrographs. As a result, few of these systems exist. The need for autonomous transient characterization led to the construction of at least three ``intermediate-resolution'' and one low-resolution robotic spectrographs within the last decade. LCO's Global Telescope (LCOGT) Network contains one of these systems, the Folded Low Order whYte-pupil Double-dispersed Spectrograph (FLOYDS). FLOYDS is a pair of intermediate-resolution (R = 400 - 700) spectrographs built for the classification of supernovae and other variable sources~\citep{2011AAS...21813203S}. FLOYDS is installed on the 2~m telescopes on Haleakala (FTN) and in Australia (FTS). SPRAT, the Spectrograph for the Rapid Acquisition of Transients, is also an intermediate-resolution system (R$\sim$350) designed to rapidly follow up and classify transients down to magnitudes of $V\approx20$~\citep{2014SPIE.9147E..8HP}. SPRAT is mounted at the 2~m robotic Liverpool telescope in La Palma, Spain.

The Spectral Energy Distribution Machine (SEDM) was built on the founding concepts behind FLOYDS and SPRAT but with a low-resolution (R$\sim$100) integral field unit (IFU) spectrograph. SEDM was built to follow-up dim transients detected by PTF, down to an $r$-band magnitude of $\approx$20.5 in an hour-long exposure with a signal-to-noise ratio (SNR) of 5~\citep{2018PASP..130c5003B}. SEDM is located on the Palomar 60-inch telescope (P60) and has been in operation since April 2016. SEDM will shortly be joined by SEDMv2 on the Kitt Peak 2.1~m telescope; that system is expected to be able to observe a similar number of targets as SEDM, about 15 per night, but down to deeper limiting magnitudes. 

\subsection{ArgusSpec's Fast-followup Concept}
With existing resources, it is impossible to follow up a significant fraction of the expected detections from the Evryscopes, Pathfinder, and the Argus Array with low-latency spectroscopic observations. Attempting to build a new large spectroscopic follow-up telescope required to routinely reach the required depths would take many years, with a budget in the many \$10M's range. As an alternative approach for flare science, we built the low-cost ArgusSpec system designed to characterize only the brightest transient detections from these systems ($m_V<13$). ArgusSpec serves as a prototype system for a possible ArgusSpec Array, which would be capable of rapidly following up the bright and dim transients detected across the sky by the full Argus system by multiplexing many individual ArgusSpecs to observe large numbers of bright targets per night, or fewer dim targets through coordinated observations. 

ArgusSpec is a fully autonomous, rapid-slew spectroscopic follow-up system designed to characterize M-dwarf flares detected by the Argus Pathfinder. The instrument is built from a rapid-slew mount, a 16-in.\@ Ritchey-Chr\'etien (RC) telescope, and a low-resolution spectrograph (R$\sim$150). We note that the current version of ArgusSpec's spectrograph-camera combination has quite a non-optimal final pixel scale, which reduces system SNR. ArgusSpec was developed from easily available, off-the-shelf components and parts in stock in our lab. With automated acquisition and operational robustness demonstrated by this initial prototype, future versions of the system will have greatly improved SNR enabled by custom optics with more appropriate pixel scaling.

The ArgusSpec approach was inspired by SEDM, taking advantage of low-resolution spectroscopy to broadly characterize transients across the sky. ArgusSpec uses a software suite built from the ground up for low-latency observing that enables the system to go from receiving an observation request to reduced spectra within tens of seconds. Using alerts from its companion all-sky telescopes, ArgusSpec will build a spectroscopic library of flares and superflares. The instrument was deployed to the Pisgah Astronomical Research Institute (PARI) in Western North Carolina on 1 Dec 2022; commissioning was completed on 22 Feb 2023. 

In this paper, we present the hardware, design, and initial on-sky performance of ArgusSpec. In Section~\ref{sec:hardware} we describe the ArgusSpec hardware and optical design, in Section~\ref{sec:software} we detail the custom python control software suite and the data reduction pipeline, and in Section~\ref{sec:testing} we discuss the instrument's on-sky performance and show some initial data taken by ArgusSpec during testing. In Section~\ref{sec:summary} we discuss the potential for building an ArgusSpec Array and summarize.

\section{Instrument Design and Hardware\label{sec:hardware}}
The chief requirements for ArgusSpec are a 20$\sigma$ limiting magnitude of $m_V\approx13$ at 30 s, at 500 nm, with a spectral resolution of $R\approx150$. The limiting magnitude requirement is based on the number of flare targets produced each night by Argus Pathfinder, as a function of target brightness (the superflares targeted by ArgusSpec outshine the stars by many magnitudes at peak, greatly increasing the target numbers achievable even at these relatively bright limits). The resolution requirement is based on separating the bright emission lines of large M-dwarf flares from the continuum blackbody emission, temperature fitting of hot superflares up to 40,000 K, and (with improved UV transmission) using the method described in \cite{2013ApJS..207...15K} to measure the ratio of blackbody to pseudo-continuum line emission at the Balmer jump. The evolution rates of large flares (e.g. \citealt{2018ApJ...860L..30H}) require rapid target acquisition, with the goal of being on-target within a minute of event onset. The 30 s exposure-time requirement is based on the typical peak evolution rate of the flares \citep{2020ApJ...902..115H}. ArgusSpec was designed to a $\rm\$35k$ hardware budget.

\begin{figure}
\centering
  \includegraphics[width=.48\textwidth]{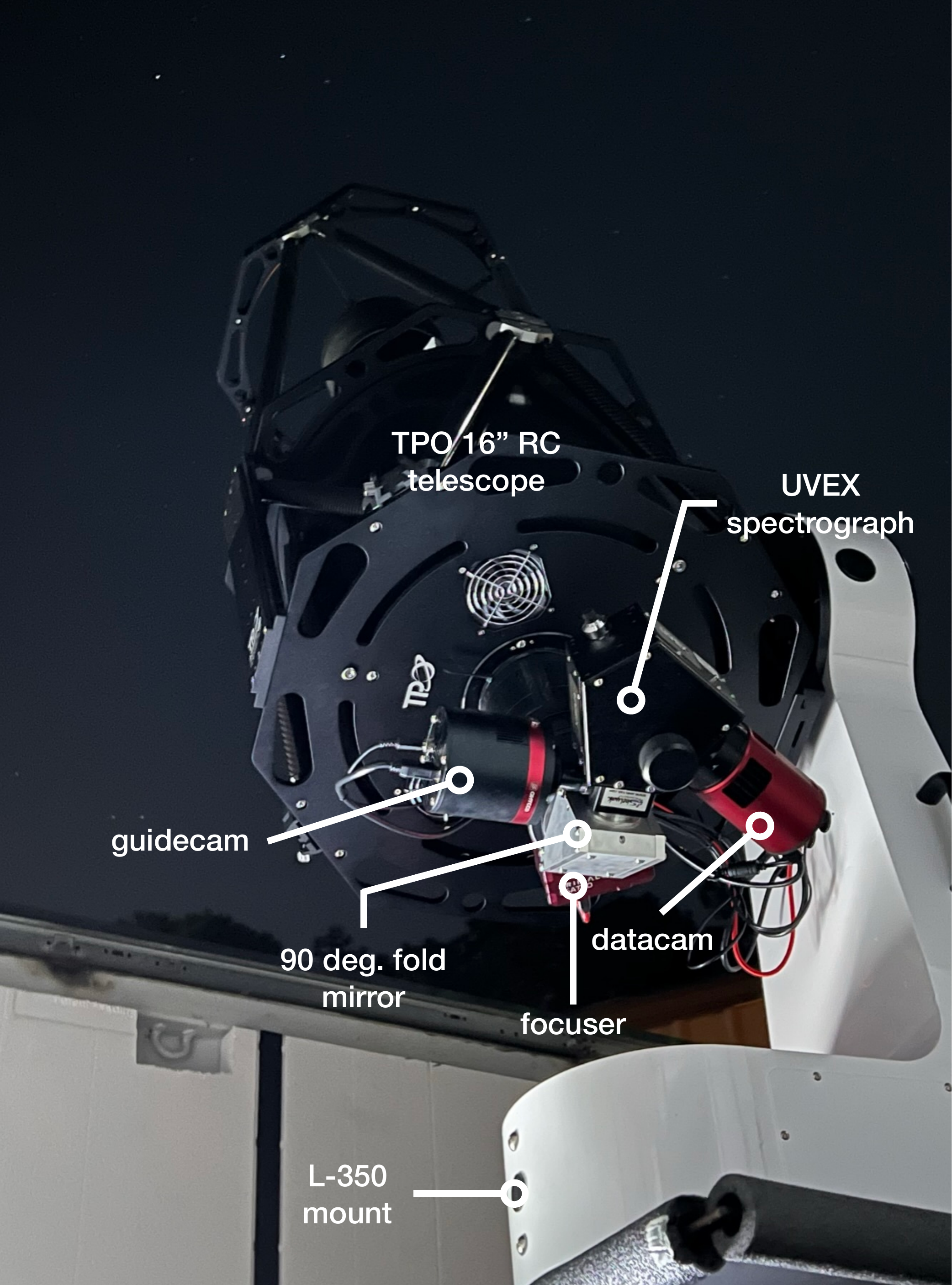}
  \caption{The assembled ArgusSpec optical train with key components labeled. \label{fig:optical_assm}}
\end{figure}

\begin{figure}
\centering
  \includegraphics[width=.48\textwidth]{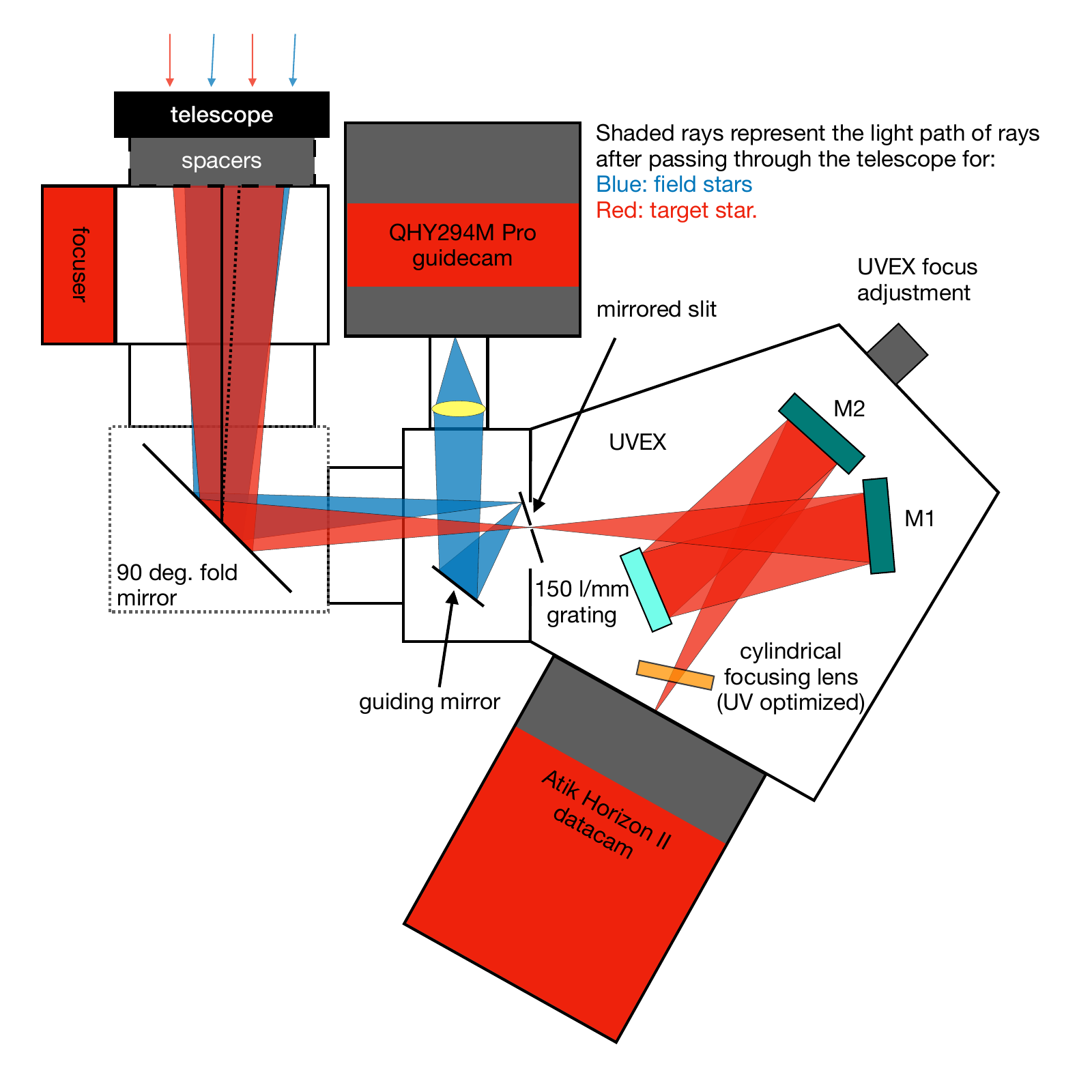}
  \caption{A diagram of the path of light through the ArgusSpec optical train. Light from field stars is reflected off of the mirrored surface of the slit and observed by the guidecam. Light from stars positioned on the slit will pass through the spectrograph and a spectrum will be captured by the datacam. The folding mirror orientation and the path of light coming from the telescope have been rotated into the plane of the page for clarity. The figure is not to scale. \label{fig:light_path}}
\end{figure}

The ArgusSpec optical train is made up almost entirely of affordable, consumer off-the-shelf (COTS) astronomical equipment. The telescope is a TPO 16-inch (f/8) Ritchey-Chr\'etien, made by GSO optics.\footnote{\url{http://www.gs-telescope.com}} The telescope beam passes through a PrimaLuceLab\footnote{\url{https://www.primalucelab.us}} ESATTO 3 Inch Robotic Microfocuser, which has a resolution of .04 microns per step. The beam then passes through a custom-made 90 deg. fold mirror ($>$ 85\% throughput over our observing wavelengths). This is the only custom-made part in the optical train, which is required for the telescope back focus to be mechanically compatible with the mount arm height. The light is then focused onto a reflective 50~$\mu$m slit and is dispersed by the Ultra-Violet Explorer (UVEX) spectrograph. The slit and the UVEX spectrograph are both produced by Sheylak Instruments.\footnote{\url{https://www.shelyak.com}} The slit surface is mirrored and is observed with a QHY\footnote{\url{https://www.qhyccd.com}} 294M Pro 11.7 MPix CMOS camera for target acquisition and guiding. The telescope plus optical train is mounted on a PlaneWave\footnote{\url{https://planewave.com}} L-350 direct drive mount. The assembled optical train can be seen in Figure~\ref{fig:optical_assm} and a diagram of the light path through the optical train is shown in Figure~\ref{fig:light_path}. A calibration box containing a Helium lamp is located inside the ArgusSpec container and is controlled remotely via an NPS.

Here we summarize the ArgusSpec hardware designs and specifications; details on the control system and data reduction pipeline are provided in Section~\ref{sec:software}. Table~\ref{table:instrument_summary} summarizes the system specifications. 

\subsection{Enclosure}
\begin{figure}
\centering
  \includegraphics[width=.48\textwidth]{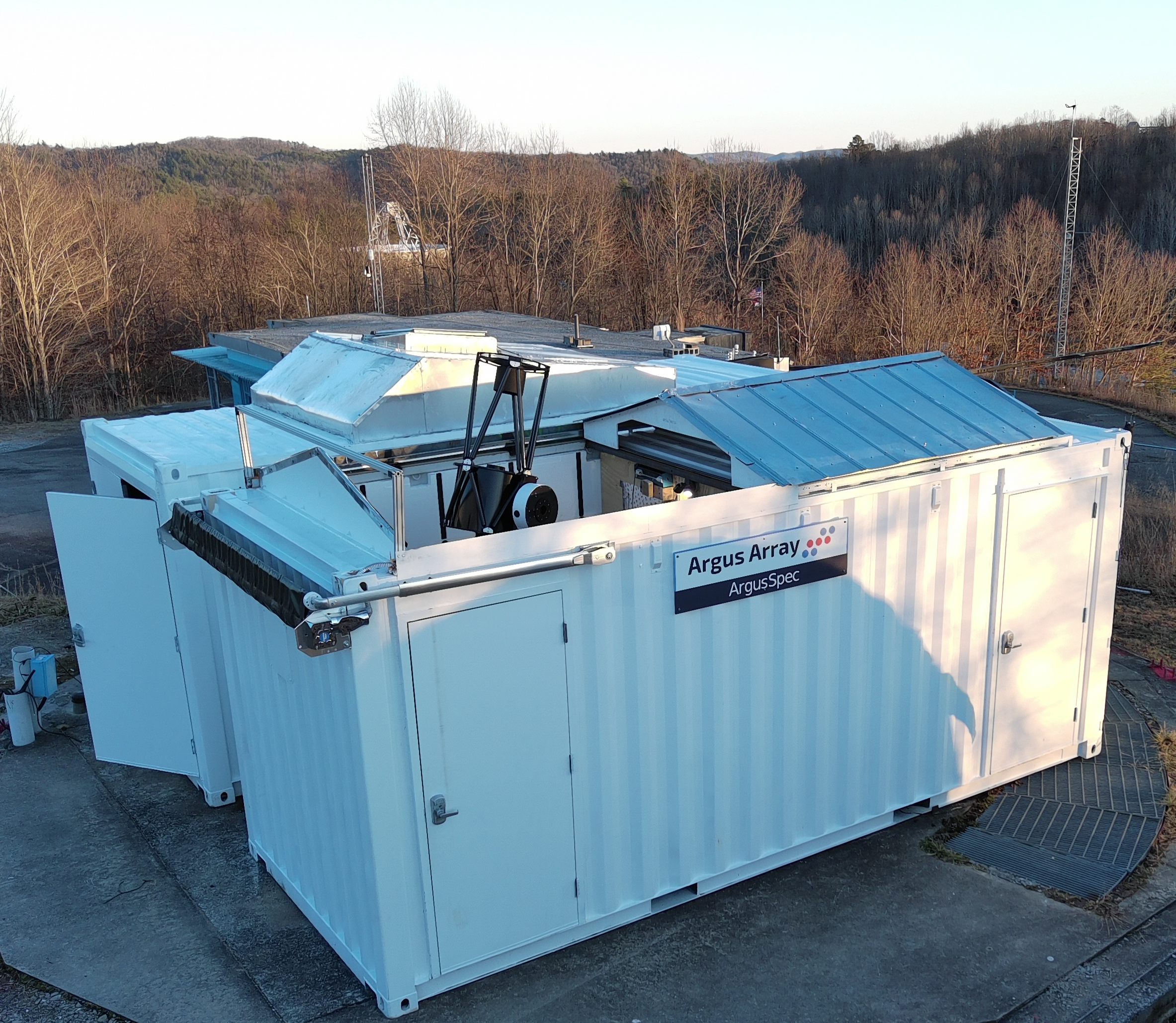}
  \caption{The ArgusSpec (front) and Argus Pathfinder (back) instruments deployed at PARI. The instruments were assembled in two shipping containers, with half of the ArgusSpec container dedicated to housing their shared, environmentally-controlled server room (right door). \label{fig:containers}}
\end{figure}

The ArgusSpec and Pathfinder instruments (Figure~\ref{fig:containers}) were designed to be assembled inside modified shipping containers. This choice was made to reduce budgetary costs for housing the instruments and to reduce the time required to perform on-site assembly. The shipping containers also supply a robust, built-in method of transportation and allow for rapid deployments which only requires moving the containers to a new location. Two major modifications, a roll-off roof and a $\sim$5 ft tall steel pier, were necessary to allow ArgusSpec to observe from inside the shipping container. 

\begin{table}
	\caption{Hardware specifications for the ArgusSpec system.\label{table:instrument_summary}}
	\renewcommand{\arraystretch}{.75}
    	\begin{tabular*}{\columnwidth}{p{0.42\columnwidth}p{0.52\columnwidth}}
    	 \hline\hline
	 \qquad\qquad Telescope& \\
	 Equipment \dotfill & TPO 16~in.\@ RC\\
	 Aperture \dotfill & 16 inches (406 mm)\\
	 Obstruction \dotfill & 7.5 inches (190 mm)\\
	\hline
	 \qquad\qquad Spectrograph& \\
	 Equipment \dotfill & Shelyak Ultra-Violet Explorer (UVEX)\\
	 Optical design \dotfill & Cross Czerny-Turner\\
	 Grating \dotfill & 150 l/mm blazed echelette grating (blazed at 500~nm)\\
	 Optical Resolution \dotfill & R = 156 @ 500~nm (150 l/mm grating)\\
	 Bandpass \dotfill & 370~nm - 750~nm\\
	 Efficiency \dotfill & 35\% at 550~nm (including camera QE, not including slit losses, modeled)\\
	 Slit \dotfill & 50~$\mu$m ($\sim$2.7$''$)\\
	\hline
	 \qquad\qquad Data Camera& \\
	 Equipment \dotfill & Atik Horizon II Mono\\
	 Sensor specs \dotfill & 16 MPix CMOS; 4644 $\times$ 3506 px \\
	 Read noise \dotfill & 1e$^-$ (gain setting 30)\\
	 Pixel size \dotfill & 3.8~$\mu$m\\
	 Dark current \dotfill & 0.016 e$^-$/pix/s (at -10°C)\\
	 Readout speed \dotfill & .125 s/image (full frame)\\
	 Gain \dotfill & 0.16 e$^-$/ADU (gain setting 30)\\
	 Standard exposure length \dotfill & 30 s\\
	 \hline
	 \qquad\qquad Guide Camera& \\
	 Equipment \dotfill & QHY294M Pro\\
	 Sensor specs \dotfill & 11.7 MPix CMOS; 4164 $\times$ 2796 px (full frame not used for guiding)\\
	 Read noise \dotfill & 1.6 e$^-$ (high gain mode)\\
	 FoV \dotfill & 10\arcmin $\times$ 8.5\arcmin~(limited by the mirrored slit surface size, not sensor size)\\
	 Pixel size \dotfill & 4.63~$\mu$m\\
	 Dark current \dotfill & 0.005 e$^-$/pix/s (at -10°C)\\
	 Readout speed \dotfill & .0625 s/image (full frame)\\
	 Standard exposure length \dotfill & 3 - 7 s\\
	 \hline
	 \qquad\qquad Mount& \\
	 Equipment \dotfill & PlaneWave L-350 Direct Drive Mount\\
	 Max speed \dotfill & 50 deg/s\\
	 Pointing accuracy \dotfill & $<$10$''$~RMS \\
	 Pointing precision \dotfill & 2$''$~at sidereal velocity\\
	 Tracking accuracy \dotfill & 0.3$''$~over a 5-minute period at sidereal velocity\\
	 Load capacity \dotfill & 100 lbs (45 kg)\\
	 \hline
	\end{tabular*}
\end{table}

\subsection{Telescope Selection\label{sub:telescope}}
The selection of the telescope to be used for ArgusSpec was based on several criteria:

\begin{itemize}
  \item Minimal refractive optics, to optimize UV throughput.
  \item Large enough aperture to reach scientifically-motivated limiting magnitudes.
  \item Backfocus and weight compatible with the chosen mount.
  \item Collimation robust to seasonal and nightly temperature changes.
  \item f/\# compatible with the chosen spectrograph.
  \item Physical size fits within the allowed footprint inside a shipping container.
\end{itemize}

We found the TPO 16~in.\@ (f/8) RC telescope to be the best option for meeting all of the above criteria. RC telescopes have no refractive optics and therefore block no UV light. With a 16~in.\@ aperture, we estimate using standard signal-to-noise (SNR) calculations the instrument can reach 20 $\sigma$ limiting magnitudes of $m_V\approx 13$ at 30 s cadence, with a resolution of R$\sim$150, through the spectrograph. This is deep enough to see several large flares per night from Pathfinder. CAD modeling showed that a larger telescope would not fit within the required footprint of the shipping container. 

The TPO telescope has a carbon-fiber truss tube design, which has demonstrated the ability to maintain collimation and focus relatively well over dramatic temperature changes occurring on hour-to-hour and day-to-day timescales. In some cases, we see collimation drift during temperature changes of 30-40 degrees Fahrenheit, but the telescope returns to acceptable alignment when the extreme weather subsides. A seasonal-adjustment of collimation is all that is necessary to maintain good optical performance. Focus for the system holds throughout entire observing nights, and frequently over periods of days and weeks, needing only minor adjustments. As further motivation for selecting this optical design, the UVEX spectrograph was optimized for use with f/8 RC telescopes. 

The only problem requiring custom work was making the back focus compatible with the L-350 mount. Once the telescope is mounted and balanced, there is roughly 14~in.\@ of clear space between the support base of the mount and the back of the telescope when pointed at zenith. However, the back focus of the TPO 16~in.\@ is 260 mm ($\sim$10 inches) and the UVEX is $\sim$6 inches tall. This resulted in the spectrograph making contact with the mount when pointed at high elevations. To resolve this problem, we designed and built a custom 90 deg.\@ fold mirror (Figure~\ref{fig:optical_assm}). To support the instrumentation, we built the fold mirror with a sturdy housing to also serve as the mechanical support for the rest of the optical train. The mirror has $>$85\% throughput over our observing wavelengths.

The optical performance of the telescope during testing has been acceptable for the deployment site, which often has few arcsecond seeing. Through the guiding camera, we achieve image quality better than 3.0$''$ full width at half maximum (FWHM) on clear nights in median seeing. Guiding camera image quality is dominated by the surface quality of the reflective slit and the guidecam optics provided by the guiding module. Imaging performance was not a priority for optimization in the current ArgusSpec system, and there is room for significant improvement in subsequent hardware iterations. Further details on the acquisition and guide camera hardware are detailed in Section~\ref{sub:guiding_hardware}.

\subsection{Target Acquisition and Guiding Hardware\label{sub:guiding_hardware}}
Target acquisition and guiding are done by the same optical system, which is referred to as the ``guidecam'' going forward. Light coming through the optical train can take one of two paths after passing through the fold mirror: 1) light from stars located on the slit will be dispersed and focused by the spectrograph, to be observed by the data acquisition camera; 2) light from field stars will be reflected by the mirrored slit onto the guiding mirror, and then passes through a lens group and is focused onto the guidecam camera's sensor. The guiding system optics have an f/\# = 1 so that the image is formed on the camera's sensor without magnification.

The guiding camera is a QHY294M Pro. It has an 11.7 MPix 4/3" CMOS sensor, with 4.63~$\mu$m pixels and fast readout. It has a low dark current at 0.005 e$^-$/pix/s (at -10°C) and a low read noise of 1.6 e$^-$. The sensor is much larger than the usable surface area of the mirrored slit, which has a size of 9.5 $\times$ 7.5~mm. Therefore, we use a region-of-interest (ROI) from the guiding camera during slit acquisition and guiding to decrease image transfer and processing times. The usable ROI FoV of the guiding system is approximately 10$'$ $\times$ 8.5$'$. The guidecam images have distortions near the edges of the field, caused by the uneven surface of the slit, so we often crop the ROI further to reduce these effects. The guide camera was selected from parts on hand to rapidly develop and prototype the ArgusSpec system. Due to this, the field of view and plate scale of the guiding system are suboptimal and require further improvement. A more suitable pairing of guide camera and optical train will be evaluated for future revisions to the system.

This setup allows ArgusSpec to slew to a target, observe the pointing of the telescope, create an astrometric solution for that field, autonomously align the star onto the slit, and then guide on the star for the duration of the observation. The limited FoV of the guiding camera does make targeting some fields difficult, as there may not be enough stars to create an astrometric solution. Given the small FoV of the off-the-shelf guide camera, solutions are only reliable within 25° of the galactic plane (80\% success rate with current hardware). The large majority of flare targets will be found in this region, and ArgusSpec has met its science requirements with this performance. Extension to higher galactic latitudes simply requires a larger guidecam FoV. A future, improved version of the ArgusSpec system might use a larger mirrored slit and a slightly faster telescope to expand the FoV of the guiding system. Further details on the automated slit acquisition and guiding software are provided in Section~\ref{sec:software}.

\subsection{Spectrograph and Data Camera\label{sub:spectro_hardware}}
The UVEX is a Cross Czerny-Turner spectrograph optimized for observations from the near UV ($<$400~nm) to the near-infrared (1~$\mu$m). We chose to use the lowest-resolution grating offered with the UVEX, a 150 l/mm blazed echelette grating. The grating is blazed at 500~nm, optimized for throughput in the visible bandpass. We use a 50~$\mu$m slit ($\sim$2.7$''$), chosen to optimize light throughput for the expected seeing conditions at PARI.

The ``datacam'' is an Atik Horizon II Mono, with 3.8~$\mu$m pixels. The sensor is a 16 MPix, 4/3" CMOS chip with extremely low read noise (1 e$^-$ at gain 30) and low dark current (0.016 e$^-$/pix/s at -10°C). We chose the Horizon II for its low read noise characteristics, which could allow for the coadding of sequential spectra to reach fainter limiting magnitudes. The Horizon II also met the back focus specifications of the UVEX, which has a limit of 13 mm. 

The UVEX and datacam have a combined estimated efficiency of 35\% at 550~nm. This estimate does not include potential slit losses from slit alignment errors, tracking errors, or bad seeing. We chose the widest slit available for the spectrograph to help offset some of these effects and optimize total light throughput.  An advantage of the Alt-Az configuration of the mount is that the angle of the slit relative to the parallactic will always remain constant. Given careful alignment during commissioning, ArgusSpec minimizes the effects of differential atmospheric refraction with standard parallactic-aligned observations across the sky.

\begin{figure*}
  \includegraphics[width=.99\textwidth]{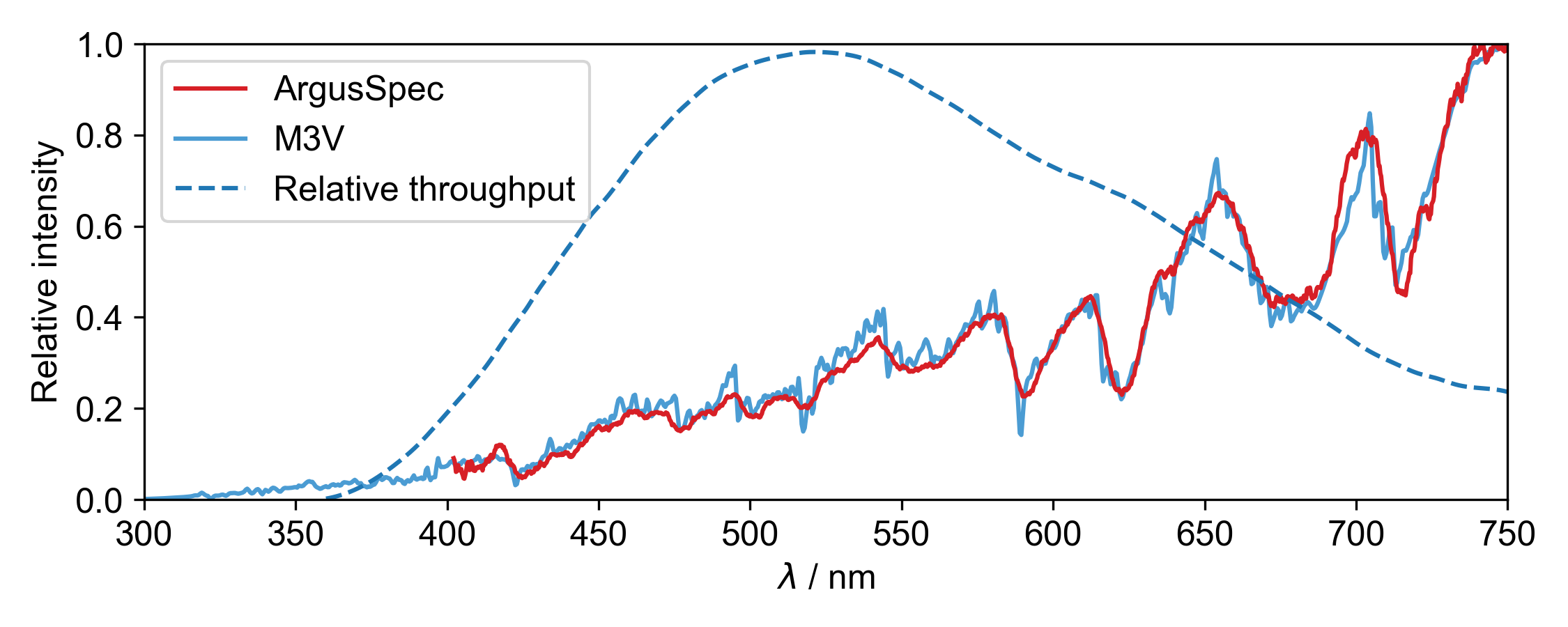}
  \caption{A 30~s ArgusSpec exposure of an $m_V=11.7$ M3V star. The red line shows the reduced ArgusSpec data, the solid blue line shows the Pickles Atlas~\citep{1998PASP..110..863P} catalog spectrum for an M3V star, and the dotted blue line shows the system throughput. The flux values have been normalized.\label{fig:throughput}}
\end{figure*}

We found the usable bandpass of the system to be 370 - 750~nm; below 370~nm system throughput rapidly falls to zero (Figure~\ref{fig:throughput}). This is limited primarily by the camera window, which will be replaced in future system iterations. Above 750~nm, SNR falls off rapidly. The signal loss in that region is mostly dominated by camera QE drop-off, not throughput from the UVEX.

The resolution of the system is dominated by the slit width. The slit is 50~$\mu$m (or $\sim$13 datacam pixels), and the UVEX contains no internal focus reducer, so the projection of the slit on the datacam is the physical size of the slit. The wavelength range 400 - 750~nm covers $\sim$1400 pixels on the sensor. Therefore, there are approximately 110 resolution elements across the 400 - 750~nm wavelength range. This gives a resolution element of 3.2~nm, or an R$\sim$150 at 500~nm. We note that the current spectrograph-camera combination pixel scale results in reduced system SNR as a result of ArgusSpec being assembled from readily available off-the-shelf components and parts in stock in our lab. Future ArgusSpec versions will have improved SNR through a more optimal spectrograph-camera pairing.

\subsection{Rapid-Slew Mount Overview\label{sub:mount_hardware}}
The PlaneWave L-350 mount was chosen for its ability to support the required heavy load of the ArgusSpec optical train (90~lbs), its ability for precise and accurate pointing (2$''$ pointing precision at sidereal velocity, with $<$10$''$~RMS pointing accuracy), and its fast slew speed of 50 deg/s. The mount is capable of operating in alt-az mode, which made designing the pier support system inside the ArgusSpec container considerably easier and allows simple parallactic alignment. An existing software application allows for Pythonic interfacing for control, real-time status reports, and creating pointing models. The direct-drive motors provide smooth movement which helps reduce slit losses during data acquisition. 

\subsection{Construction of the Roll-off Roof and Support Pier for the Shipping Container\label{sub:system_construction} }
To serve as a robust, moveable observatory, we started with a standard shipping container and made significant modifications. This allowed the system to be assembled and tested in Chapel Hill and then relocated to the final observatory site. We determined this approach would allow for the easiest method for rapid deployment, with the possibility of re-deploying the instrument somewhere else in the future. It also was the most cost-effective option compared to a traditional astronomical dome. For ArgusSpec, these modifications required cutting a hole in the top of a container, building a powered roll-off roof, and constructing a steel support pier to support the instrument above the opening in the roof. 

\subsubsection{Roll-off Roof}
A 7.5~ft.\@ $\times$ 7.5~ft.\@  square hole was cut from the roof of the container, to allow ArgusSpec to move freely without risk of hitting the container when mounted on the pier. This is effectively the widest hole possible in the shipping container without removing material from the external support frame. The size of the hole needed was estimated using CAD modeling and was one of the deciding factors in determining the size of the telescope we could use. This hole was positioned as far north in the container as possible, to maximize the space left for the service module, at the other end of the same container. 

We built a roll-off roof to weatherproof the telescope and container when the system is not in use. The frame of the roll-off roof is approximately 7.5~ft.\@ $\times$ 8~ft.\@ $\times$ 2 ft.\@ in size. It was framed with 1~in.~$\times$ 1~in.\@ aluminum T-slot rails, except for the support base of the frame which is 2~in.~$\times$ 2~in., and follows a standard A-frame design. The framing is reinforced with a layer of 1/4~in.\@ pine plywood on the top, front, and back surfaces. The front and back plywood cutouts were made on a Computerized Numerical Control (CNC) machine. The cutouts were then primed and painted before being attached to the roof. Weather-proofing was completed by screwing galvanized steel roofing panels to the plywood on top of the roof. 

The roof is mounted onto garage-door tracks, which are welded onto the container, using rollers mounted to the framing. The roof is opened and closed using a standard garage door opener, which is controlled with an Arduino. The same Arduino monitors and provides updates to the control system on the roof state, by monitoring two limit switches on either end of the roof's range of motion.

\subsubsection{Emergency Rain Cover}
In the event of a roof failure, ArgusSpec has a backup emergency rain cover. The emergency system is an unmodified motorized tarp system for a dump truck, fitted onto the front of the ArgusSpec container. The backup cover is controlled by the roof Arduino, which monitors for status updates from the control system and in the event of a roof failure will deploy the emergency tarp. This system can be seen on the front of the container in Figure~\ref{fig:containers}. The backup system is powered by a standard 12V car battery.

\subsubsection{Support Pier}
The support pier follows a simple design, made of three 6~in.\@ square tube legs, that support a 1/4~in.\@ steel plate. Mounted on top of the pier is a 1/2~in.\@ thick mounting plate for the L-350, which has 6$\times$ 3/8~in.\@ tapped holes for securing the mount. Between the top plate of the pier and the mounting plate, we placed a sheet of rubber gasket, to help reduce the passage of vibrations from the container to the mount.

\begin{figure}
\centering
  \includegraphics[width=.48\textwidth]{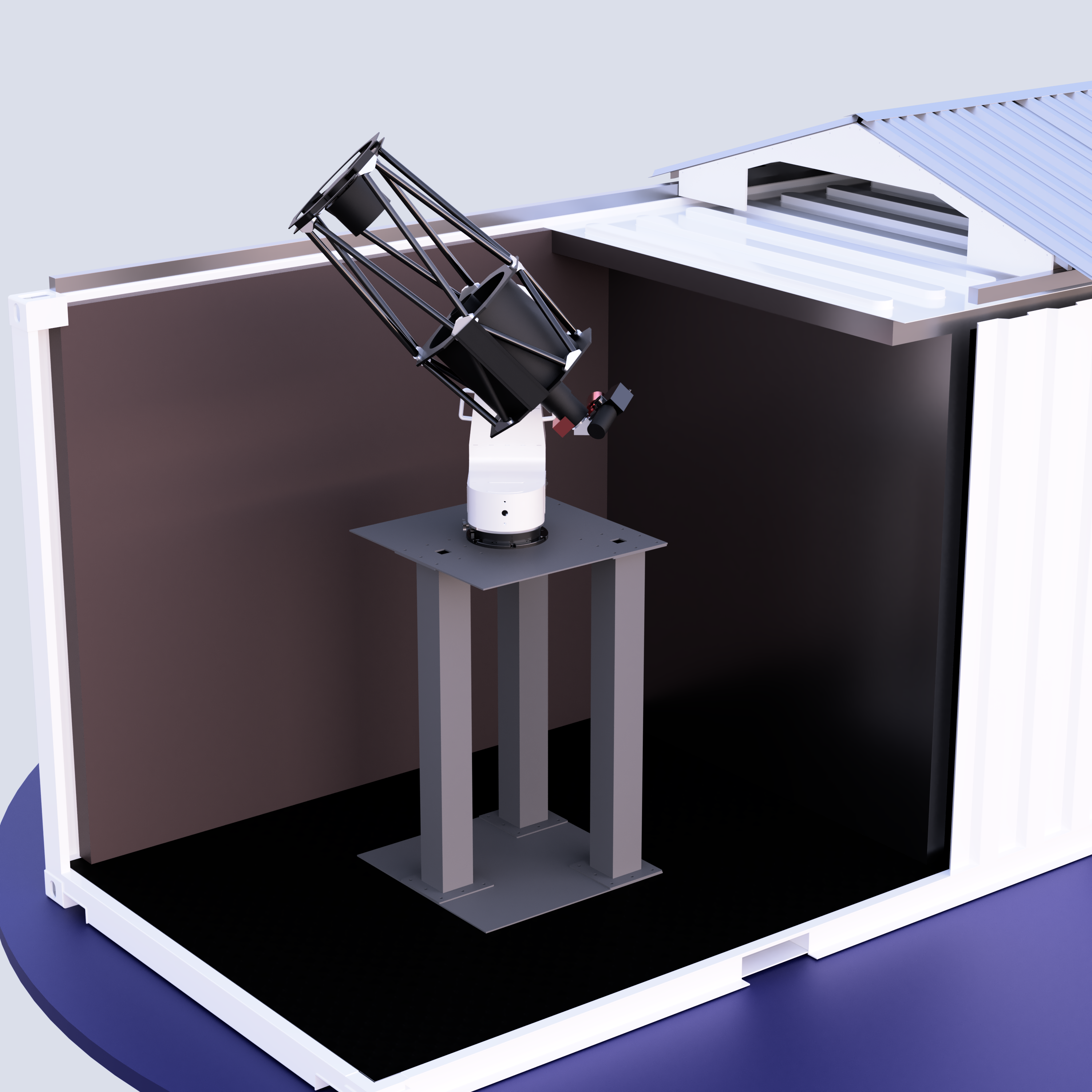}
  \caption{A cutaway rendering of ArgusSpec inside its shipping container, displaying the steel support pier. The pier is 4.75~ft.\@ tall and was designed to reduce vibrations through the use of vibration-dampening rubber gaskets between the L-350 mounting plate and the steel structure. The pier is secured into the framing of the container with several dozen 3/8~in.\@ lag bolts. When the mount is in its park position, the roof can close and create a weather-resistant environment by pushing against an end wall attached to the end of the container (not shown).\label{fig:pier}}
\end{figure}

The pier structure is attached to the frame of the container with 3/8~in.\@ lag bolts that are screwed through the wood flooring and into the metal framing. The pier is 4.75~ft.\@ tall, which allows ArgusSpec to view to an altitude of 10 deg.\@ or less in the East/West directions. In the North/South directions, the FoV is blocked by the roof and the end cap up to an altitude of $\sim$20 deg. A cutaway rendering of the telescope mounted onto the pier is shown in Figure~\ref{fig:pier}.

\subsection{Wiring and Electronics\label{sub:wiring}}
The wiring architecture is simple, with most control signals and data handling being done with a single control computer and a couple of Arduinos. ArgusSpec shares power, network storage, weather monitoring, and internet with Pathfinder.

\subsubsection{Power Distribution and Control}
The ArgusSpec and Pathfinder containers were outfitted with standard electrical circuitry. The power feed from PARI into the electrical box in the service module is clean power supplied from a universal power supply (UPS). Inside the ArgusSpec container is a set of four standard 120V wall outlets and two ground fault circuit interrupter (GFCI) wall outlets. 

The power for any electronics that might need power cycling is provided by a Digital Loggers Network Power Switch (NPS), which allows computer-controlled switching of individual power ports. It has eight switchable ports, five of which are currently in use for the system, powering: the control computer, mount, guidecam, datacam, and lamp box. The lamp box is a simple single-tube calibration box, containing a helium lamp, which is toggled on and off with the NPS for calibrations. 

The roof opener is powered by the GFCI-protected outlets, which have an emergency stop button inside the container. The emergency backup tarp is powered by a 12V car battery, which is kept fully charged using a standard trickle charger. The Arduino in charge of monitoring and controlling the roof and emergency tarp is powered by both the wall outlet and the car battery. In the case of total power loss for the system, the roof Arduino can still deploy the emergency backup tarp if necessary. The roof and the backup tarp are triggered by running their open/close connections through an Arduino relay shield, which is triggered by a signal from the control computer.

Occasionally the cameras suffer hardware or software failures that require power cycling to reset. We found that cycling just the power connection via the NPS was not sufficient to reset the cameras to a usable state; the cameras required manual unplugging and re-plugging the USB connections to cut the 5V bus power. To solve this problem, we use USB cables with switchable power bus connections and control those connections remotely using an Arduino with a relay shield. 

\subsubsection{Data and Control Signal Distribution}
System control and data reduction are handled by a sealed, fan-less OnLogic Helix 600 Intel Comet Lake rugged computer. The cameras, mount, and supporting Arduinos are connected to the control computer via USB connections. The control computer supports all of these connections without the need for a USB hub. The control computer runs the control and data reduction daemons (Section~\ref{sec:software}), storing all resulting data products onto the shared Pathfinder network-attached storage (NAS) in the service module. The site is directly connected to the North Carolina Research and Education Network, with negligible downtime and a 100 Mbps allocation. The system is designed to enter a safe state upon loss of communication with the internet, should a failure occur. Additionally, all data processing needed for normal operation of ArgusSpec, including generating astrometric solutions, is performed on-site.

\subsubsection{Weather Monitoring}
Weather is monitored by a Diffraction Limited\footnote{\url{https://diffractionlimited.com}} Boltwood Cloud Sensor III. The Cloud Sensor is powered by the Pathfinder NPS so that it can be reset in case of hardware or software failure. It is connected to the network via WiFi and provides updates on cloud cover, rain, wind, sky brightness, and several other environmental measurements. The Cloud Sensor readings are regularly monitored by the control daemons to make sure rain, cloud cover, wind, daylight, humidity, and ambient temperature sensor readings are appropriate before opening and during observing. 

\subsubsection{Webcam Monitoring}
\begin{figure}
\centering
  \includegraphics[width=.48\textwidth]{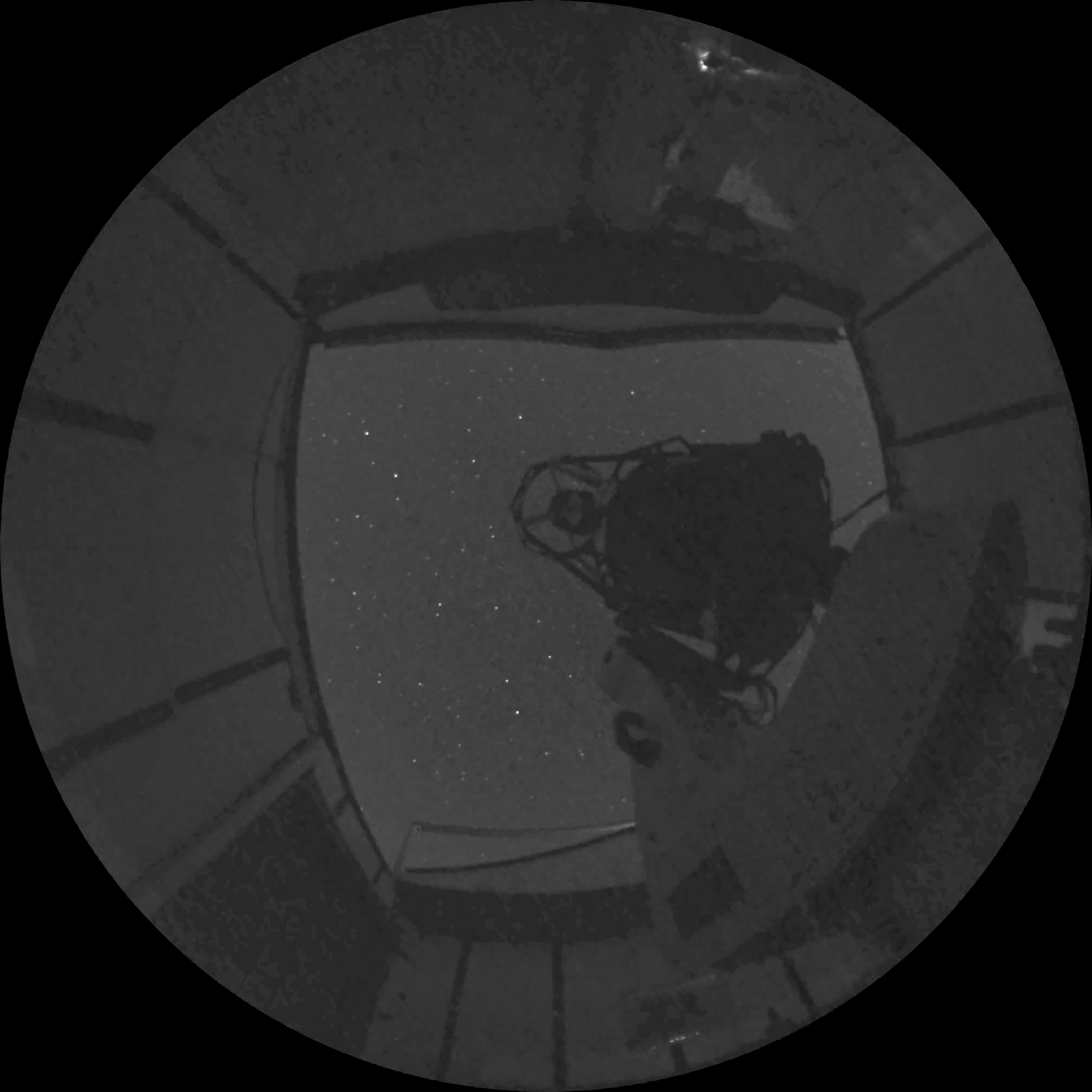}
  \caption{An image taken from the ArgusSpec webcam during observations. \label{fig:webcam}}
\end{figure}

An AXIS M3067-P network camera is installed onto the pier. This allows for the remote monitoring of the system, to diagnose any potential issues. The webcam gives a fisheye view inside the container, allowing the state of the roof, mount, optical train, and telescope to be checked in one view. A sample image from the webcam taken during observations is shown in Figure~\ref{fig:webcam}.

\subsubsection{Lightning Protection}
Electrical storms at PARI are quite common. To prevent equipment damage the containers are grounded to an earth grid via an external grounding cable with a gauge specified for lightning protection. All electronics are powered through the NPS, which has built-in surge protection. The telescope pier and the 12V battery, which powers the emergency backup tarp, are also grounded to the container's common ground. So far the system has experienced no electrical surges or equipment damage from electrical storms.

\section{Custom Software Suite: System Control and Data Reduction \label{sec:software}}
ArgusSpec is controlled by a custom Python framework running on the control computer. The framework is built around the use of several daemons. The daemons are set up as state machines, each of which controls a range of tasks related to a specific piece of hardware or functionality. They communicate via JSON-based TCP/IP sockets. Contingencies are built into the system to protect against unexpected software failures. For example, the roof and backup tarp are controlled by an Arduino that monitors and communicates with the roof daemon, and in the event of an unexpected software failure, deploys the emergency cover. In this section, we discuss the roles of each of the daemons and describe key software features, such as automated slit acquisition and guiding. We also provide details on the data reduction pipeline. 

\subsection{System Control}
\begin{figure}
\centering
  \includegraphics[width=.48\textwidth]{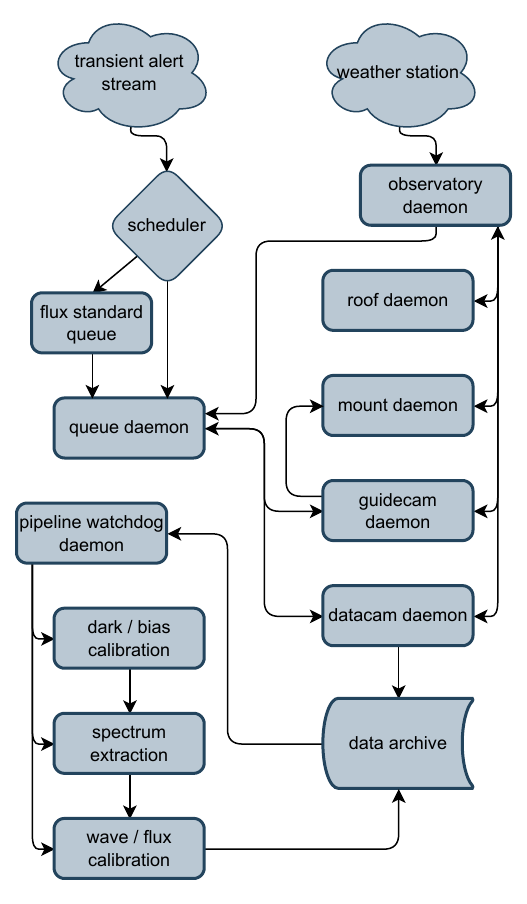}
  \caption{A block diagram of the ArgusSpec control daemons and data reduction processes. Double-headed arrows represent cross-communication between two daemons. \label{fig:daemon_block}}
\end{figure}

To reduce the risk of hardware loss due to software failure, critical daemons controlling the roof, mount, and overall observatory state were designed to be simple. They have limited functionality and have been tested extensively to ensure the chance of critical failure is minimal. More complex daemons, such as the guidecam daemon, are not critical in observatory control in the case of a failure while the system is open. Figure~\ref{fig:daemon_block} shows an overview of how all of the daemons and processes are connected. In the following sections, we describe each of the daemons and their functionality.

\subsubsection{Roof Daemon}
The roof daemon is in charge of tracking the state of the roof, communicating that state to other daemons, and issuing open/close commands to the roof-controlling Arduino. The Arduino uses two limit switches mounted inside the container, at both ends of the opening in the roof, to report the status of the roof. The roof Arduino is constantly monitoring for pings (``heartbeats'') from the roof daemon, and in the event of a software failure, will deploy the backup tarp if it has not received a heartbeat within the last several minutes (and the roof is in an open state). 

\subsubsection{Mount Daemon}
The mount daemon is in charge of tracking the state of the fast-slew mount, receiving commands from the observatory (Section~\ref{sub:obs}) and guidecam daemons (Section~\ref{sec:guidecam}), issuing those commands to the mount, and monitoring the position of the roof to ensure the system is in a safe state before moving. This daemon communicates with the PlaneWave Interface 4 (PWI4) mount control client to issue commands to the mount. A large part of this subprocess was built around the pre-existing code base provided for the mount from PlaneWave Instruments. 

\subsubsection{Observatory Daemon: System Control}\label{sub:obs}
The observatory daemon is in charge of issuing system-wide commands for opening, closing, and taking calibrations. It monitors the weather station and sunrise and sunset times at PARI to determine when the system should perform each of those actions. In the case of borderline weather, where the weather station might suggest a rapid succession of opening and closing operations is necessary, we have built in a time check between system openings to reduce wear on moving parts. If the daemon detects that the weather station has failed it will automatically trigger observatory closure. 

\subsubsection{Datacam Daemon: Data Acquisition}
The datacam daemon controls the Atik data camera, tracks the state of the hardware, reports the hardware state and receives commands from the queuing daemon (Section~\ref{sub:queue}), and issues commands to the camera. This daemon handles the data products from the datacam by creating the headers and file names for the images and saving them to the Pathfinder NAS or local storage directories. During operation, the datacam daemon will monitor images from the camera and in the case of hardware malfunction will automatically perform a power cycle. 

\subsubsection{Guidecam Daemon: Slit Acquisition and Guiding}\label{sec:guidecam}
The guidecam daemon has two main functionalities: 1) handling, issuing commands to, and tracking the state of the QHY guide camera hardware; 2) autonomous slit acquisition and guiding during observations. The simpler of these two functionalities, handling and tracking the QHY camera hardware, is done in the same way as the datacam daemon. 

Autonomous slit acquisition and guiding are more complicated, requiring fast astrometric solutions for a continuous stream of images and coordinating the positioning of the mount in real-time. When a command is issued to the queuing daemon (Section~\ref{sub:queue}) necessary target information, such as sky coordinates and guidecam exposure time, is relayed to the guidecam daemon. Slit acquisition and guiding proceed as follows:

\begin{enumerate}
	\item The queuing daemon issues an observation request to the guidecam daemon (Section~\ref{sub:queue}).
	\item The guidecam daemon commands the mount daemon to slew to the target coordinates.
	\item Once slewing is complete, the guidecam takes an image of the star field. 
	 \item The guidecam daemon solves the star field image using a locally hosted and speed-optimized instance of \url{astrometry.net} \citep{2010AJ....139.1782L}. Improved solving speeds were achieved by constraining the astrometric solution by providing astrometry.net with current coordinates, the RA-Dec rotation angle, and guide camera pixel scale, along with pre-convolved images and custom source extraction software. Using this solution, a sky position for the target and the slit is calculated. Months of on-sky testing have shown the slit position in the guide camera's frame to be consistent within $\sim$1-2 pixels, so the slit location is calculated by transforming the known pixel location to sky coordinates via the astrometric solution.
	 \item The astrometric solution is used to calculate the offset between the target and the slit, and then jog the mount to the correct location; the high-resolution encoders ensure a reliable move at accuracies better than a tenth of an arcsecond. We have found that this one move is sufficient to bring the target into the slit in almost all our test cases.
	 \item Steps 3 and 4 are repeated for guiding. Because the telescope is an alt/az mount with no rotator (to naturally keep the slit at the parallactic angle), field rotation means that guiding cannot be easily performed from a single star. The astrometry.net solution is updated as guiding continues. ArgusSpec has performed guiding continuously for hours without issues.
\end{enumerate}

\begin{figure*}
  \includegraphics[width=.99\textwidth]{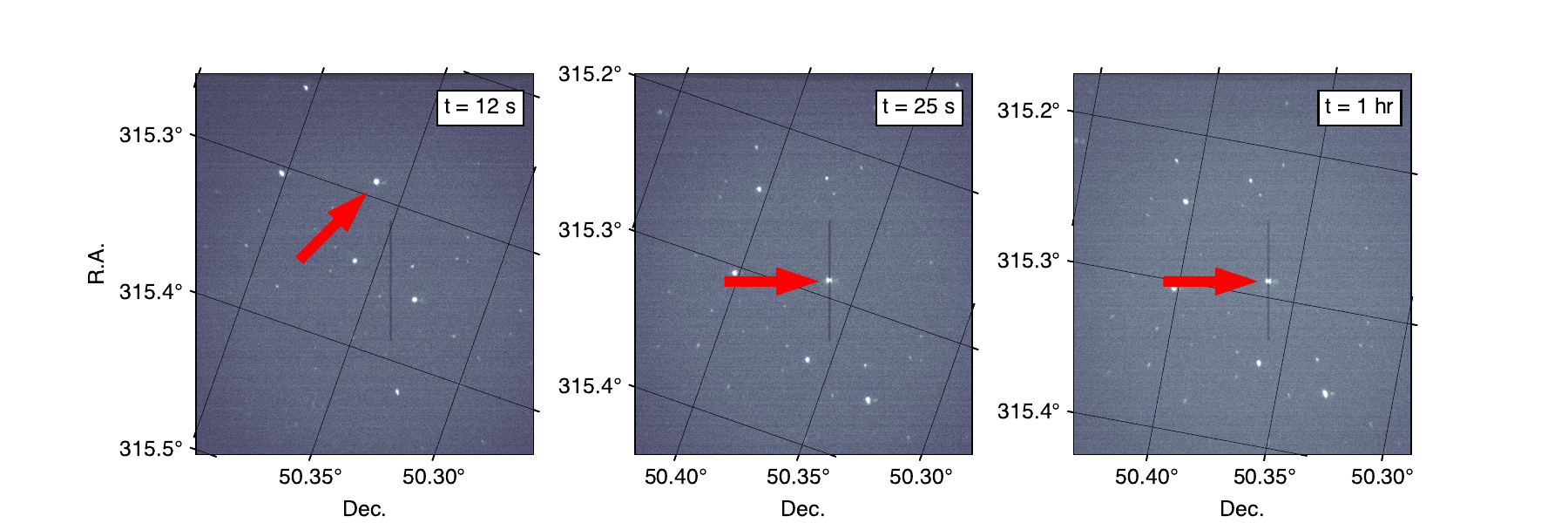}
  \caption{A set of sample guidecam images taken at the start and end of observations for V1331 Cygni, a young stellar object. Each image was taken with an exposure length of 7 s. The time denoted in the upper-right corner of the plot shows the delay from the initial observation request being sent to the queuing daemon to the start of the guidecam exposure. The first image on the left is the initial star field taken by the guidecam after slewing to the target field. The second image shows the position of the target and the slit after the first field solve and mount jog, with this initial move being enough to place the target star on the slit. The last image on the right shows a third guidecam image taken one hour after the second image, to demonstrate tracking stability over extended observations. Over a single night of observations, it was found that ArgusSpec had an RMS tracking stability of 1.1 arcseconds in right ascension and 1.2 arcseconds in declination.\label{fig:guideimages}}
\end{figure*}

For a large fraction of targets, astrometry.net solves within 4 seconds and guiding proceeds as described above. Given an average initial slew time of $\sim$7 seconds and a guidecam exposure time of 5 seconds, ArgusSpec can begin guiding within 16 seconds of receiving an alert. A set of sample guidecam images taken at the beginning of observations of V1331 Cygni are shown in Figure~\ref{fig:guideimages}. 

Astrometric solutions do become more difficult in less-dense sky regions given the guider's small FoV, meaning there are locations on the sky that do not solve, and therefore ArgusSpec can't observe targets. This is uncommon but does occur for some subset of targets. A more common occurrence is that a field will solve only some fraction of the time. This often happens because the target star is a key source in solving the astrometry, but when placed on the slit is no longer detected as a source by the extraction code. 

In these instances, slit acquisition and guiding are still possible. Once a single astrometric solution has been found, the first jog of the mount almost always places the target star somewhere within the slit. In the following guide image, we measure flux values at the expected location on the slit, and if there is a significant deviation from the background, we assume the target is placed reasonably well for observations. In subsequent images, if no solution is found, we monitor both sides of the slit looking for changing or unbalanced flux values. If one side begins to display a higher flux, we jog the mount to counteract this motion. In this mode, we have been able to observe targets for over two hours without astrometric solutions. 

For targets where the daemon does not find a single astrometric solution after several attempts, the system will begin semi-randomly jogging the mount to nearby sky positions in an attempt to find a more densely populated field. This is a last resort backup method, which results in much slower targeting times if the system can find a solution at all. 

\subsubsection{Queuing Daemon: Target Handling}\label{sub:queue}
The queuing daemon is responsible for handling observation requests from external sources. The queue daemon is also in charge of creating and managing directories for data storage, which it communicates (along with other key information) with the two camera daemons at the start of each observation. When a target observation request is received by the queuing daemon it: 

\begin{enumerate}
	\item Checks the state of all of the other daemons to make sure the system is in a safe state for observations.
	\item Issues a command to the guidecam daemon to begin acquisition and guiding.
	\item Monitors the status of the guidecam, and when the guidecam acquires the target, issues the command for the datacam to begin taking exposures. 
	\item When observations are finished, it makes sure all daemons return to their waiting state and are prepared for future observations.
\end{enumerate}

The queuing daemon is also responsible for reporting the status of the observatory to external sources waiting to issue observations. It will add to the queue any commands sent to it if the system is already observing another target, or deny the requests if the hardware is not in a safe state to observe.

In cases where immediate follow-up is required, but the system is busy observing another target, you can tell the daemon to stop all other observations and observe the high-priority target instead. The queuing system for ArgusSpec has not yet required target priority handling or optimization. That challenge will be handled separately, outside of the ArgusSpec code base in future work done alongside Argus-HDPS~\citep{2022SPIE12189E..10C} code base for Pathfinder and Argus.

\subsection{Data Reduction}
The data reduction pipeline is based on the Evryscope-SOAR flare reduction pipeline (Corbett et al. 2023, in prep.), with several minor modifications. We briefly describe it here for completeness. We also discuss the removal of two noise patterns contained in our datacam images.

\subsubsection{Datacam Noise Removal}
A hardware bug in the datacam causes every image to contain a checkered grid of 16 $\times$ 16-pixel squares that are offset from one another by a constant value. We believe this pattern is some uncorrected processing that was used for the color version of the Atik Horizon II. We remove this pattern for all images before processing by mapping, and then subtracting, the constant offset in each of the squares. The Atik camera also displays slight noise from electrical interference in the x-direction across the chip. We measure the noise for a given y-value on the chip by measuring the median value across 1000 pixels located in the same row, but outside of the footprint of the spectra produced by the spectrograph. This median value is subtracted from each row in the image. Each image is then bias and dark subtracted before spectral extraction and reduction. During the commissioning phase of the project, it was found that turning the flat fielding of the detector on and off had a negligible effect on the final output spectra of the system, suggesting that the pattern removal was successful.

\subsubsection{Spectra Reduction}
At the beginning and end of each observing night, we collect He calibration lamp spectra for wavelength calibrations. However, the calibration box is mounted in the container and therefore is not observable for different airmass configurations of the system. To account for the shifting of the spectrum as a function of telescope altitude, due to mechanical flexure of the system from a changing gravity vector on the optical train, a simple x-position linear offset is applied to the lamp wavelength solutions. The offset is calculated from the x-position of the zeroth order peak. Final science-quality spectra are further wavelength corrected by fitting known emission and absorption lines in the spectra. 

We use an automated extraction routine to extract the target and background sky spectra. Each image is binned along the spectral axis by a factor of 18. Each bin with a significant SNR is fit with a Gaussian plus a constant x-position linear offset, obtaining a FWHM and y-location value for each x-axis bin. The average of the FWHM values is stored as the FWHM of the spectrum. We run a random sample consensus \citep[RANSAC]{fischler1981random} fit, using the locations of the Gaussians as inputs, to find the best-fit polynomial for the location of the spectrum in the image. This fit is used to produce a 2D Gaussian extraction image, which is then multiplied by the calibrated science image and then summed along the spectral axis to produce a 1D spectrum. The background is measured in the same way, but we offset the 2D extraction image by $\pm$7, 10, 15, and 19 $\times$ the FWHM of the spectrum along the spectral axis before extracting the 1D sky spectra. The sky spectra are sigma-clipped, averaged, and then subtracted from the science spectrum.

Spectra are flux calibrated using ESO spectrophotometric standards \citep{1994PASP..106..566H}. We take several images (5-20, depending on the standard's brightness) and average them together to improve SNR. We calculate a flux scaling based on the ratio between observed and catalog continuum emission from the standard star. Both the observed standard spectrum and the catalog values are masked to omit absorption and emission features and smoothed using a Savitsky-Golay filter \citep{1964AnaCh..36.1627S}. Science spectra are then multiplied by this flux ratio and normalized.

Data reduction takes an average of $\sim$14 s per spectrum, so the pipeline can be run in real-time for any observing cadence slower than that. We built a custom pipeline daemon that monitors and automatically reduces new images being taken by the datacam. The pipeline daemon watches for new observations of standards and automatically detects which standard observations to use for current reductions based on proximity to the current target's airmass. This automated pipeline daemon does not produce science-quality data, as the wavelength calibrations are often imprecise. This is due to the fitting used to estimate the x-pixel offset of the spectrum as a function of airmass; the fit can vary due to slight changes in the shape of the zeroth order peak of the spectrum as the telescope tracks. In final science reductions, we adjust the wavelength solutions by fitting to emission and atmospheric lines. This real-time reduction pipeline is used to monitor the observations from the instrument quickly to look for events that should be further investigated.

\section{On-sky Testing \label{sec:testing}}
In this section, we present some preliminary data taken by ArgusSpec during a series of on-sky tests that have been performed since commissioning. With Pathfinder still undergoing commissioning, all of the on-sky testing for ArgusSpec has been done by using known active flare stars as targets to validate the data reduction pipeline; or by simulating an event stream to test its targeting speed. Future work will contain science results from ArgusSpec's follow-up of rapid transient detections from the Pathfinder array. 

\subsection{Time to Acquisition}
An important metric for ArgusSpec's performance is the time to acquisition, or the amount of time between issuing an observation request and acquiring the target star on the slit. For flares, and many other rapid transients, the key rise and fall times can be on the order of tens of seconds. 

We tested ArgusSpec's on-sky time to acquisition through a series of repeated observations. To begin each observation, we commanded the system to slew to a random altitude and azimuth. This simulates the system having finished a previous observation at a random sky position. We then commanded the system to observe a preset target, 1RXS J155702.5-195037 \citep{2000IAUC.7432....3V}, and measure the time it takes for the system to acquire the target in the slit. This process was repeated over 150 times. We chose to have the system observe a preset target, instead of a random sky-field, during each iteration because we knew the target location would be on a star and not an empty patch of sky. This allows the system to use both of its methods for determining whether the target star had been positioned on the slit (Section~\ref{sec:guidecam}). 

\begin{figure}
\centering
  \includegraphics[width=.48\textwidth]{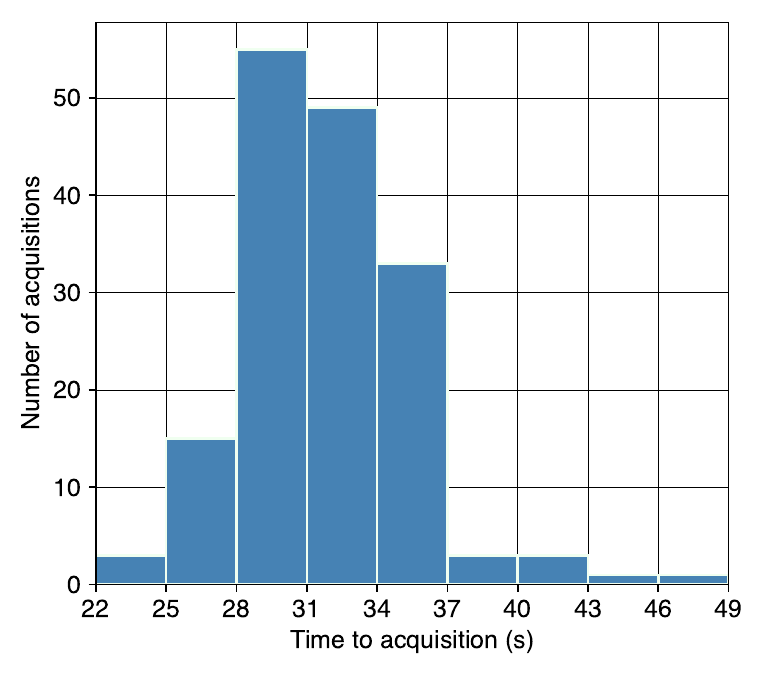}
  \caption{A histogram showing the results of an experiment to measure ArgusSpec's target acquisition time from a random starting altitude/azimuth position. The bin size along the x-axis is 3 s. We commanded ArgusSpec to a random sky position and once settled, asked it to observe a preset target. This process was repeated over 150 times and during each observation the system recorded the time it took to slew, solve the initial field, and then place the target star onto the slit via one or more mount jogs. This test used a guidecam exposure time of 3 s.  The average and median acquisition times were 31.5 s and 32.5 s, respectively.  \label{fig:acquisition_histo}}
\end{figure}

The results of this experiment are shown in Figure~\ref{fig:acquisition_histo}. The system had a median time to acquisition of 32.5 s, using 3 s guidecam exposures. Roughly a third of this time is spent slewing the mount, which can be reduced in the future by increasing the mount speed. Currently, the mount operates in the 20 deg/s mode, but we plan to increase this to 30 deg/s or more. Guidecam image processing currently takes $\sim$5 s to find an astrometric solution and jog the mount, before beginning another exposure. To process the images faster, a custom astrometric solver would be required. It is important to note that these results are representative of fields where the guidecam daemon readily finds astrometric solutions. In less-densely populated sky regions, time to acquisition will likely be slower, or in some cases, targeting the object may not be possible. This is a limitation of the current guide camera which will be upgraded in subsequent hardware evolution.

Given Argus Pathfinder operating in its rapid 1-second cadence with Argus-HDPS processing and sending out alerts within 1 second, ArgusSpec will be able to follow-up flares during these key times with an average acquisition time of around 30-40 s from the onset of detectability of the flare’s impulsive phase (and potentially faster; software accounts for at least half of this latency). Guidecam observations, providing multiband photometry compared to Pathfinder, will begin within 20 s of the flare onset. For the brightest superflares (e.g. \cite{2018ApJ...860L..30H}) these capabilities will enable the following of the flare temperature throughout the flare peak emission.

\subsection{Limiting Magnitude}

\begin{figure}
    \centering
    \includegraphics[width=.48\textwidth]{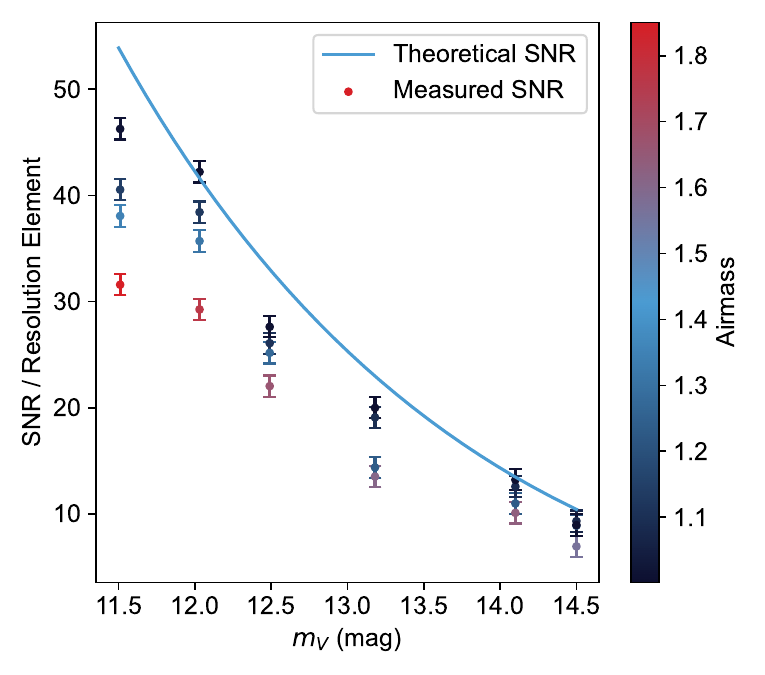}
    \caption{ArgusSpec SNR measurements for different target V band magnitudes. Each measurement was taken multiple times throughout an observing night to obtain measurements at different airmasses. These values are plotted alongside the system's theoretical performance, which was calculated assuming 35\% throughput from the spectrograph, a 50\% cut from atmospheric transmission, and an additional 70\% efficiency from the optics for a total throughput of 13\%. The theoretical performance does not account for slit losses.}
    \label{fig:snr_plot}
\end{figure}

A series of 30-second exposure images were taken 10 minutes at a time throughout a single night for a set of stars with known V band magnitudes \citep{2000A&A...355L..27H, 2020yCat.1350....0G}. These images were then reduced and had their pixel positions mapped to their respective wavelengths. For each image, a photometric aperture was centered at the 500 nm wavelength of each spectrum. To measure background counts, a grid of background apertures was placed far enough above and below the spectra to avoid measuring any signal from the source. System performance was calculated using the standard SNR equation,
\begin{equation}
    \frac{S}{N} =\ \frac{N_{s}}{\sqrt{N_{s}+N_{B}+n_{pix}t_{exp}\phi_{D}+n_{pix}{\sigma}^{2}}},
    \label{SNR_eq}
\end{equation}
where $N_{s}$ is the number of electrons collected in the photometric aperture centered at 500 nm, $N_{B}$ is the mean number of electrons collected per aperture across the grid of background apertures, $n_{pix}$ is the number of pixels per aperture, $t_{exp}$ is the exposure time of the image, and $\phi_{D}$ and $\sigma$ are the dark current and spatially averaged readout noise of the CMOS sensor \citep{2022zndo...6825092B}. The readout noise, $\sigma$, was determined by taking the standard deviation across all background apertures and $\phi_{D}$ was taken from the datacam's datasheet. For each series of exposures a mean SNR was calculated. These values can be found in Table \ref{tbl:snr_table} and are compared against the theoretical system performance in Figure \ref{fig:snr_plot}. 

Currently, exposure times for the datacam are set manually. In the future, this process will be automated in the following way. For a target of unknown brightness, a test exposure will be taken for both cameras, and the SNR of the image will be calculated while looking for saturated pixels within the target. From these measurements, the initial exposure time may be halved or doubled if needed.

\subsection{Slit Losses}
Slit losses were quantified by taking a series of observations of a set of stars with V band magnitudes ranging from 9 to 14, both inside and outside the slit. A nearby star was used to normalize the flux between both sets of measurements. The flux received while inside the slit was divided by the flux received outside to calculate slit loss. It was found that ArgusSpec experiences a weighted average slit loss of 42\%, including instrumental and atmospheric effects. Clearly, subsequent hardware iterations will require this to be improved.

To evaluate any wavelength-dependent effects due to slit misalignment with the current system, we purposefully misaligned a star to 0.1 slit widths from the slit edge. Under the current parallactic-angle alignment of the system, this maximal misalignment scenario resulted in a maximum of 20\% deviation from a center-aligned spectrum, measured at both extreme edges of the ArgusSpec wavelength coverage. This performance is already within the science requirements, but in normal operations, the slit alignment is factors-of-several better, and wavelength-dependent effects will be concomitantly lower.

\begin{table}[]
    \caption{ArgusSpec SNR performance for various target magnitudes.}
    \medskip{}
    \begin{tabular}{lll} \hline \hline 
    Star                     & $m_V$ & SNR $\pm 1.0$   \\ \hline 
    TYC 2727-1012-1          & 11.51            & $46.2$ \\
    TYC 2727-284-1           & 12.03            & $42.2$ \\
    TYC 3198-1784-1          & 12.49            & $27.6$ \\
    TYC 3198-952-1           & 13.18            & $20.0$ \\
    GPM 331.516461+31.736605 & 14.1             & $13.3$ \\
    GPM 331.767570+32.183741 & 14.5             & $9.3$\\
    \hline
    \end{tabular}
    \label{tbl:snr_table}
\end{table}

\subsection{AD Leonis Staring Campaign}
AD Leonis (AD Leo) is a relatively bright ($m_V = 9.3$) M3.5V active flare star located in the northern sky. With its high probability of undergoing at least a small flare throughout just a few observing nights, it has served as a good test bed for the ArgusSpec system. ArgusSpec has observed AD Leo for several nights continuously since deployment, in a series of staring campaigns which have provided data for testing the data reduction pipeline. 

\begin{figure*}[h]
  \includegraphics[width=.99\textwidth]{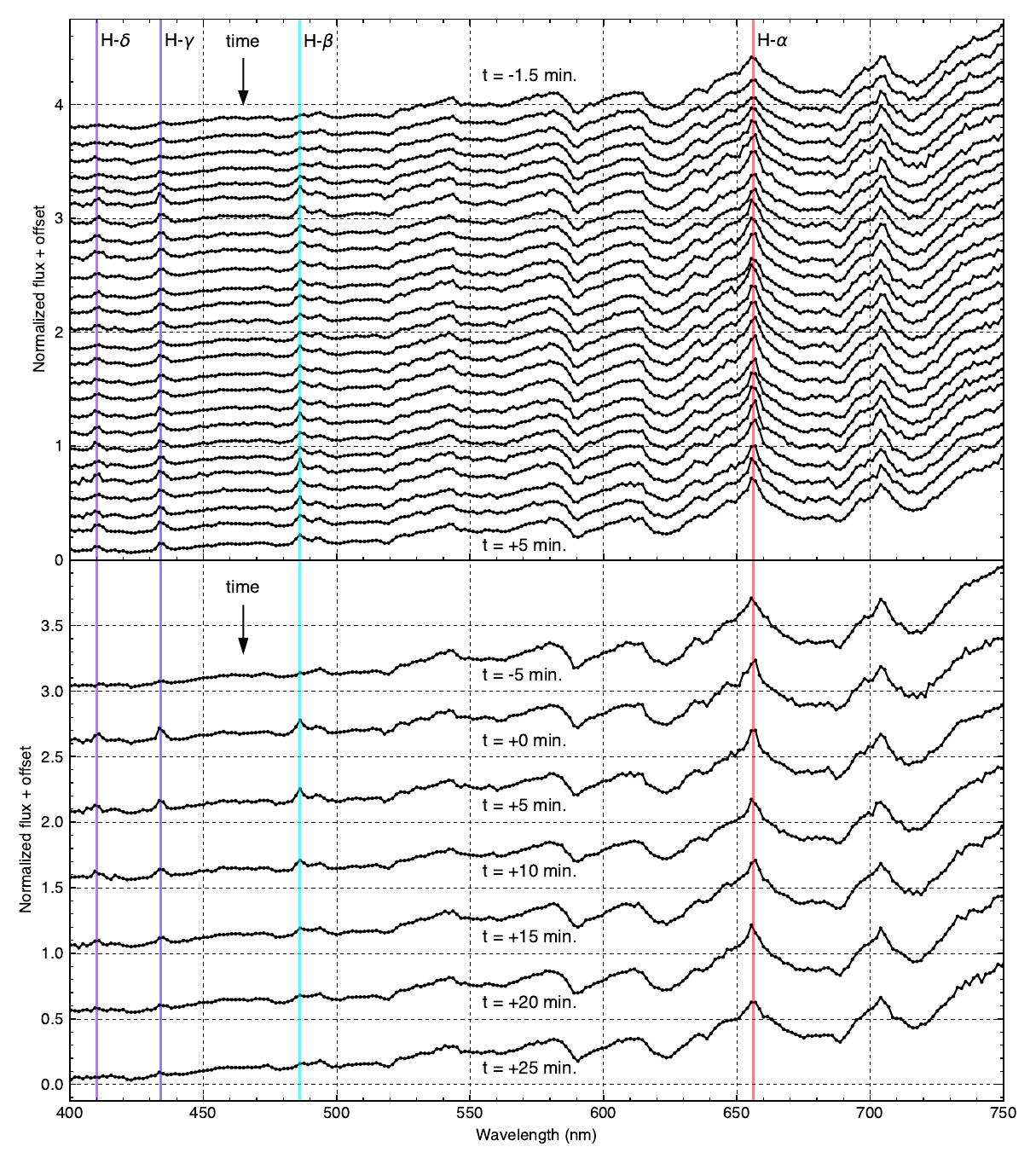}
  \caption{Spectra of a small flare captured by ArgusSpec during observations of AD Leo on the night of 13 Apr 2023. Each spectra shown was taken with a 15 s exposure time. The top plot shows every spectrum taken by ArgusSpec over a $\sim$6.5 minute period, starting approximately 1.5 minutes before the peak of H-$\beta$ by the flare. The bottom plot shows the longer-timescale behavior of the flare, showing every 20th spectra taken by the datacam beginning approximately 5 minutes before the peak of the flare. The spectra have been binned down to match the optical resolution of the system. Balmer spectral lines are marked on the plot with vertical solid lines that are colored according to their wavelengths and labeled in the top panel of the plot. \label{fig:adleo_flare_spectra}}
\end{figure*}

Prior to beginning the campaign, a set of exposure times were taken during quiescence. A 15-second exposure time was chosen as it allowed a high enough signal to noise to be scientifically viable while allowing the demonstration of a higher cadence mode. The surrounding star field also allows for consistent astrometric solutions and therefore long-term tracking of the target without interruption. Throughout the staring campaigns, we have observed a few small flares with the largest of these occurring on the morning of 13 Apr 2023. Figure~\ref{fig:adleo_flare_spectra} shows the flux-calibrated spectra taken by ArgusSpec during the flare.

\begin{figure}[h]
  \includegraphics[width=.48\textwidth]{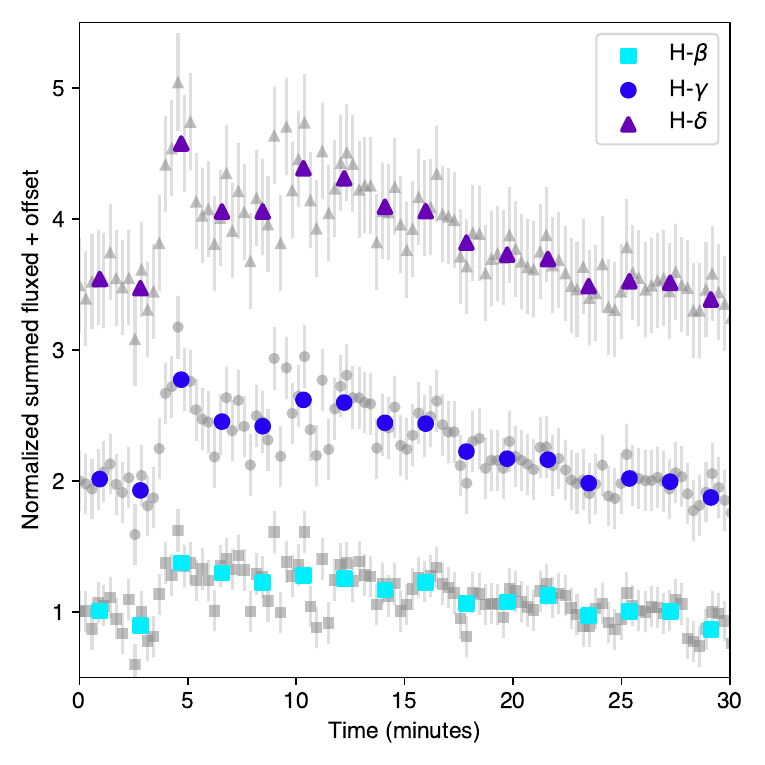}
  \caption{Light curves of the higher-order Balmer emission lines from the small flare emitted by AD Leonis and observed by ArgusSpec on 13 Apr 2023. The light curves were generated by summing the flux contained within the FWHM of each emission line and then normalizing to their median pre-peak value. The gray points show the raw data points and uncertainties and the colored points show the data binned down by a factor of 6 along the x-axis.  \label{fig:line_evolution}}
\end{figure}

The light curves for the highest SNR Balmer emission lines, H-$\beta$, H-$\gamma$, and H-$\delta$, are shown in Figure~\ref{fig:line_evolution}. These light curves were generated by summing the flux contained within the FHWM of each emission line and then normalizing the points to the median of their pre-peak values. The emission lines generally follow the fast rise, exponential decay profile expected for flares and show the expected divergent behavior with the continuum components, consistent with \cite{2013ApJS..207...15K}. Each of the light curves displays multiple peaks, making this a complex flare. The initial peak is likely due to prompt emission, while the second peak is more gradual and is similar to cooler emission due to reheating in overlying loops described previously in \cite{2016ApJ...820...95K}. The overall flare structure displayed by the emission lines is similar to ``late-phase'' solar flares \citep{Woods:2011} and the ``peak-bump" TESS flares described in \cite{2022ApJ...926..204H}.

During autonomous operation, once ArgusSpec receives an alert, it will slew to the target star and begin taking data for 30 minutes before the system becomes available to observe the next flare target. To characterize the quiescent state of the star, it will be revisited a week following its initial observation.

\subsection{System Reliability}
The custom control software suite has proven to be extremely robust. Since instrument commissioning and completion of the control software, the system has operated fully autonomously for dozens of observing nights with no software failures. The data reduction pipeline reduces data in real-time up to observing cadences of 15 s. System hardware has also proven to be robust; the roof has maintained functionality despite enduring harsh mountainous conditions which include: surviving winds over 60 mph, torrential rains, and extreme temperature fluctuations. The emergency backup tarp, while functional, has not required deployment since installation and testing. The mount pier has remained well-aligned; we have seen no drift in targeting or guiding. The system has operated in temperatures below freezing with no failures. Since deployment, the telescope has not required re-collimation and imaging and spectroscopic performance has been consistent.

\section{Summary \label{sec:summary}}
In this paper, we have discussed the construction of ArgusSpec, a new autonomous follow-up instrument. We have demonstrated that the system is capable of fully robotic operations and have characterized its on-sky performance. Now that ArgusSpec is fully operational, it will soon begin characterizing transient detections made by the Argus Pathfinder Array. ArgusSpec will undergo upgrading to improve SNR at faint magnitudes.

\subsection{Instrument Overview}
ArgusSpec is a low-resolution, low-cost, fast-response, autonomous spectroscopic follow-up system designed to characterize flares detected by the Argus Pathfinder. ArgusSpec's custom automated target acquisition routine allows it to begin autonomous observations of a target star within an average of 32 s from the time the observation request was received. In this paper, we described the ArgusSpec hardware, the custom Python control software, and the data reduction pipeline. We presented some initial on-sky test data, which demonstrated ArgusSpec's ability to observe targets down to a 20$\sigma$ limiting magnitude of $m_V \approx 13$ in a 30 s exposure (R$\sim$150 at 500~nm). We also presented observations of a small flare observed during a staring campaign of AD Leo. The entire system was designed, built, and deployed in under a year with a $\$35k budget$ and is ready to begin follow-up of real-time transient detections from the Pathfinder telescope. ArgusSpec was deployed to the dark sky observatory PARI in the mountains of Western North Carolina in December 2022 and commissioning of the instrument was completed in February 2023. ArgusSpec's primary science goal will be to rapidly follow up detections of flares and superflares to place tighter constraints on their blackbody continuum and effective temperature evolution. 

\subsection{An ArgusSpec Array}
Low-cost, mass-produced telescopes enable the construction of imaging telescope arrays with large effective apertures at greatly reduced cost (e.g. \citealt{2022PASP..134c5003L} and \citealt{2023PASP..135f5001O}). In spectroscopy, however, this is typically much harder: the lower per-pixel photon numbers introduce very significant detector noise contributions when coadding data from high- and moderate-resolution spectrographs, necessitating the photonic combination of incoming light onto single detectors (e.g. \citealt{2022SPIE12182E..1UA} and \citealt{2019BAAS...51g.124E}).

However, in the low-spectral-resolution transient followup regime we consider here, sky-photon noise dominates over CMOS single-electron-level detector noise for even moderate-aperture telescopes. This enables a low-cost, robust, and easily-replicable array spectrograph design: the multiplexing of simple spectrographs with low-noise detectors, with data coaddition performed after the exposure, purely in software, without experiencing significant read noise penalties. An array of dozens of these telescopes would be adaptable to the incoming transient stream, combining apertures to observe faint targets and rapidly splitting apart to cover dozens of bright targets simultaneously.

ArgusSpec is designed to serve as a prototype for such a system, although to be useful for deep transients the system will require very significantly improved per-telescope SNR. The next-generation ArgusSpec system, currently under development, is designed to demonstrate these capabilities.

\section*{Acknowledgements}
ArgusSpec was supported by an NSF AAG (AST-2009645) grant. The Argus Pathfinder was supported by NSF MSIP (AST-2034381) and by the generosity of Eric and Wendy Schmidt by recommendation of the Schmidt Futures program. Hank Corbett was supported by the National Science Foundation Graduate Research Fellowship (Grant No. DGE-1144081). WH is supported by NASA through the NASA Hubble Fellowship grant HST-HF2-51531 awarded by the Space Telescope Science Institute, which is operated by the Association of Universities for Research in Astronomy, Inc., for NASA, under contract NAS5-26555. This research made use of Astropy,\footnote{\href{http://www.astropy.org}{http://www.astropy.org}} a community-developed core Python package for Astronomy \citep{2013A&A...558A..33A, 2018AJ....156..123A, 2022ApJ...935..167A}, and SciPy,\footnote{\href{http://www.scipy.org}{http://www.scipy.org}} a core Python package for general scientific computing tasks. This work has made use of data from the European Space Agency (ESA) mission {\it Gaia} (\url{https://www.cosmos.esa.int/gaia}), processed by the {\it Gaia} Data Processing and Analysis Consortium (DPAC, \url{https://www.cosmos.esa.int/web/gaia/dpac/consortium}). Funding for the DPAC has been provided by national institutions, in particular the institutions participating in the {\it Gaia} Multilateral Agreement.

\bibliography{bibliography}{}
\bibliographystyle{aasjournal}

\end{document}